\documentclass[11pt,a4paper]{article}
\pdfoutput=1
\usepackage{a4wide}
\usepackage{amsfonts}
\usepackage{amssymb}
\usepackage{amsmath}
\usepackage{graphicx}
\usepackage{booktabs}
\usepackage{array}
\usepackage{color}
\usepackage{ulem}
\usepackage[small,bf]{caption}
\usepackage{cite}
\usepackage[usenames,dvipsnames]{xcolor}
\usepackage{subfigure}
\usepackage{verbatim}
\usepackage{dsfont}
\def\1{\mathbf{1}}
\def\3{\mathbf{3}}
\def\2{\mathbf{2}}

\def\th{\theta}

\def\f{\phi}

\newcommand{\zz}{\mathbb{Z}}

\allowdisplaybreaks[1]

\numberwithin{equation}{section}

\newcounter{mysubequation}[equation]

\definecolor{pink}{rgb}{1.,.2,.8}

\begin{document}

\begin{titlepage}

\vspace*{-15mm}
\begin{flushright}
SISSA 03/2015/FISI \\
TTP15-003 
\end{flushright}
\vspace*{0.7cm}

\begin{center}
{ \bf\LARGE Leptogenesis in an SU(5)$\, \times \,$A$_{\mathbf{5}}$\\[1mm]
Golden Ratio Flavour Model}
\\[8mm]
Julia Gehrlein$^{\, a,}$ \footnote{E-mail: \texttt{julia.gehrlein@student.kit.edu}},
Serguey T.\ Petcov$^{\, b,c,}$ \footnote{Also at:
 Institute of Nuclear Research and Nuclear Energy,
  Bulgarian Academy of Sciences, 1784 Sofia, Bulgaria.},
Martin Spinrath$^{\, a,}$ \footnote{E-mail: \texttt{martin.spinrath@kit.edu}},
Xinyi Zhang$^{\, d,}$ \footnote{E-mail: \texttt{xzhang\_phy@pku.edu.cn}}
\\[1mm]
\end{center}
\vspace*{0.50cm}
\centerline{$^{a}$ \it Institut f\"ur Theoretische Teilchenphysik, Karlsruhe Institute of Technology,}
\centerline{\it Engesserstra\ss{}e 7, D-76131 Karlsruhe, Germany}
\vspace*{0.2cm}
\centerline{$^{b}$ \it SISSA/INFN, Via Bonomea 265, I-34136 Trieste, Italy }
\vspace*{0.2cm}
\centerline{$^{c}$ \it Kavli IPMU (WPI), 
University of Tokyo, Tokyo, Japan}
\vspace*{0.2cm}
\centerline{$^{d}$ \it School of Physics and State Key Laboratory of Nuclear Physics and
Technology,}
\centerline{\it Peking University, 100871 Beijing, China}
\vspace*{1.20cm}

\begin{abstract}
\noindent
In this paper we discuss a minor modification of a previous
SU(5)$\, \times \,$A$_{5}$ flavour model which exhibits at leading
order golden ratio mixing and sum rules for the heavy and the
light neutrino masses. Although this model could predict
all mixing angles well it fails in generating a sufficient large
baryon asymmetry via the leptogenesis mechanism.
We repair this deficit here, discuss model building aspects
and give analytical estimates for the generated baryon asymmetry
before we perform a numerical parameter scan.
Our setup has only a few parameters in the lepton sector.
This leads to specific constraints and correlations between
the neutrino observables. For instance, 
we find that in the model considered  
only the neutrino mass spectrum with normal mass ordering 
and values of the lightest neutrino mass in the interval $10-18$ meV 
are compatible with the current data on the neutrino oscillation 
parameters. 
With the introduction of only one NLO operator, 
the model can accommodate successfully simultaneously 
even at $1\sigma$ level
the current data on neutrino masses, on neutrino mixing and the 
observed value of the baryon asymmetry.
\end{abstract}

\end{titlepage}
\setcounter{footnote}{0}

\section{Introduction}

The theoretical explanation for the observed neutrino oscillations and neutrino
masses requires physics beyond the Standard Model.
Furthermore the presence of Dark Matter  and the observed baryon
asymmetry of the Universe (BAU) support the need for a more
fundamental theory.
In the present article we will establish a connection between two of the above
mentioned observations and  investigate the Baryogenesis through leptogenesis
scenario \cite{Fukugita:1986hr} in an $\text{SU}(5)\times\text{A}_5$ flavour model.
The model we are going to discuss here is the first GUT A$_5$
golden ratio flavour model with successful leptogenesis to our knowledge.
This recently proposed 
model \cite{Gehrlein:2014wda} has the feature that $\theta_{12}$ is connected to
the golden ratio $\phi_g=\frac{1+\sqrt{5}}{2}$ via $\theta_{12}=\tan^{-1}\left(\frac{1}{\phi_g}\right)$.
Similar to the golden ratio (GR) type A  models in \cite{GoldenRatioA} the reactor angle
is predicted to be vanishing at leading order and the atmospheric angle to be maximal.
Hence, the neutrino mixing matrix $U_{\text{GR}}$ has the form
\begin{equation}
U_{\text{GR}}=
\begin{pmatrix}
	\sqrt{\frac{\phi_{g}}{\sqrt{5}}} & \sqrt{\frac{1}{\phi_{g}\sqrt{5}}} & 0\\
	-\sqrt{\frac{1}{2\phi_{g}\sqrt{5}}} & \sqrt{\frac{\phi_{g}}{2\sqrt{5}}} & \frac{1}{\sqrt{2}}\\
	\sqrt{\frac{1}{2\phi_{g}\sqrt{5}}} & -\sqrt{\frac{\phi_{g}}{2\sqrt{5}}} & \frac{1}{\sqrt{2}}
	\end{pmatrix}
	 P_{0} \;,
	 \label{eq:U_GR}
\end{equation}
which is given in the convention of the Particle Data Group \cite{Beringer:1900zz}
with the diagonal matrix $P_{0}$ = Diag$( \text{exp} (- \tfrac{\text{i} \alpha_{1}}{2}), \text{exp}(- \tfrac{\text{i}\alpha_{2}}{2} ),1 )$
containing the Majorana phases. Since the experimental values for the angles,
cf.~Tab.~\ref{tab:exp_parameters}, strongly disfavour $\theta_{13}$ to be vanishing
the leading order mixing angles have to be corrected to realistic values.
In \cite{Gehrlein:2014wda} we followed the approach based on Grand Unification where
the neutrino mixing angles receive corrections from the charged lepton sector.
Namely this model features SU(5) unification. Thereby we could explore the SU(5)
relation $\theta_{13}\approx \theta_C/\sqrt{2}$ from the non-standard Yukawa-coupling
relations $y_\tau/y_b = -1.5$ and $y_\mu/y_s = 6$ \cite{Antusch:2009gu,Antusch:2011qg},
and for the double ratio $(y_\mu/y_s)(y_d/y_e) = 12$ which are all in perfect agreement with experimental data. 

In addition to the corrections from the charged lepton sector renormalisation group
running effects (RGE)  have to be taken into account. Due to a neutrino mass sum rule
in both hierarchies only a certain mass range is allowed. For the inverted ordering
this implies large RGE effects for $\theta_{12}$ which rule out this ordering.

Since the light neutrino masses are generated via the type-I-seesaw mechanism
in \cite{Gehrlein:2014wda} the Baryogenesis through  leptogenesis mechanism can
be easily implemented. In this mechanism the dynamically generated lepton asymmetry
is converted into a baryon asymmetry due to sphaleron interactions. Thermal
leptogenesis can take place when the heavy RH Majorana neutrinos (and their
SUSY partners the sneutrinos) decay out-of-equilibrium in a CP and lepton-number
violating way.
Flavour effects \cite{Nardi:2006fx, Abada:2006fw,Abada:2006ea} 
(see also, e.g., 
\cite{Pascoli:2006ie,Branco:2006ce,Davidson:2008bu,Branco:2011zb})
can play an important role in thermal leptogenesis.
We set the scale at which leptogenesis takes place to be the see-saw 
scale $M_S=10^{13}$ GeV. In the model considered we have 
also $\tan \beta=30$ \cite{Gehrlein:2014wda}, and 
thus, the scale $M_S$ falls in the interval 
$10^9 (1+\tan^2\beta)$ GeV$<M_S<10^{12} (1+\tan^2\beta)$~GeV. 
For values of $M_S$ in this interval \cite{Pascoli:2006ci} 
the baryon asymmetry is produced in the two-flavour leptogenesis regime 
and we perform the analysis of baryon 
asymmetry generation in this regime.
 
We will see that the original model cannot accommodate for the observed value of the
baryon asymmetry due to the structure of the neutrino Yukawa matrix.
In fact, there would be no baryon asymmetry generated via the
leptogenesis mechanism.
In order to generate
a non-zero asymmetry we will introduce only one additional operator in the neutrino sector which
corrects the neutrino Yukawa matrix and subsequently affects the phenomenology of the model.

The paper is organized as follows: Section 2 is a short overview of the model building aspects
including the NLO operator. In section 3 we discuss the analytical results for the phenomenology
of the model. There we also describe the relevant formulas for leptogenesis. In section 4 we show
the results of a numerical parameter scan. We discuss the predictions for the mixing parameters
including the phases as well as for the sum of the neutrino masses, the observable in neutrinoless
double beta-decay, the kinematic mass $m_{\beta}$ and for the generated baryon asymmetry. In section
5 we summarise and conclude.

\begin{table}
\centering
\begin{tabular}{lcc} 
\toprule
Parameter & best-fit ($\pm 1\sigma$) & $ 3\sigma$ range\\ 
\midrule 
$\theta_{12}$ in $^{\circ}$ & $ 33.48^{+0.78}_{-0.75}$& $31.29\rightarrow 35.91$\\[0.5 pc]
$\theta_{13}$ in $^{\circ}$ & $ 8.50^{+0.20}_{-0.21}$& $7.85\rightarrow 9.10$\\[0.5 pc]
$\theta_{23}$ in $^{\circ}$ & $ 42.3^{+3.0}_{-1.6}$ & $38.2\rightarrow 53.3$\\[0.5 pc]
$\delta$ 	  in $^{\circ}$&$306^{+39}_{-70}$&$0\rightarrow 360$\\
\midrule
$\Delta m_{21}^{2}$ in $10^{-5}$~eV$^2$ & $7.50^{+0.19}_{-0.17}$ & $7.02\rightarrow 8.09$\\[0,5 pc]
$\Delta m_{31}^{2}$ in $10^{-3}$~eV$^2$ &$2.457^{+0.047}_{-0.047}$&$2.317\rightarrow 2.607$\\
\bottomrule
\end{tabular}
\caption{The best-fit values and the 3$\sigma$ ranges for the parameters in the normal ordering taken from \cite{Gonzalez-Garcia:2014bfa}.
}
\label{tab:exp_parameters}
\end{table}

\section{Model building aspects}
\label{sec:Modelbuilding}

The model we are going to discuss is based on the SU(5)$\times$A$_5$ model
proposed in \cite{Gehrlein:2014wda}. 
We only had to extend it minimally to accommodate successful
leptogenesis. The modification we are going to introduce 
has further implications for the phenomenology.
We focus first on the related model building aspects.
We briefly revise the leading order (LO) superpotential for the neutrino
sector, which is identical to the original model before we introduce the 
corrections. They are induced by an additional operator in the superpotential
which yields next-to-leading order (NLO) corrections to the Yukawa couplings
while the right-handed neutrino Majorana mass matrix remains unaffected. This
single higher order operator will generate a sufficiently
large baryon asymmetry to be in agreement with the experimental 
observations.

\subsection{The neutrino sector at LO}

We briefly summarise next the relevant parts of
the original SU(5)$\times$A$_5$ model from \cite{Gehrlein:2014wda}. 
We are not going to discuss the flavon vacuum alignment which
does not change at all and is given in the original paper.

The matter content of our model is organised in ten-dimensional representations
of SU(5), $T_i$ with $i = 1$, 2, 3, five-dimensional representations $F$,
and one-dimensional representations $N$ which transform as one-, three- and
three-dimensional representations of A$_5$ respectively, see also Tab.~\ref{tab:Fields}. 

Additionally, we had introduced in the original model the following flavons which will appear in the neutrino
sector. There are two flavons which transform as one-dimensional representations under A$_5$
\begin{equation}
 \label{eq:vev1}
 \langle \theta_2 \rangle = v_{\theta_2} \;, \quad \langle \epsilon_1 \rangle = v_{\epsilon_1} \;,
\end{equation}
one flavon in a three-dimensional representation
\begin{equation}
 \label{eq:vev3}
 \langle \f_3 \rangle = v_\f^{(3)} \left(0,0,1\right) \;,
\end{equation}
and one flavon in a five-dimensional representation
\begin{equation}
 \label{eq:vev5}
 \langle \omega  \rangle = \left(\sqrt{\tfrac{2}{3}} (v_2 + v_3), v_3, v_2, v_2, v_3\right) \;.
\end{equation}
Their charges under the shaping symmetries are given in Tab.~\ref{tab:Fields}.
Note especially that no new flavon appeared.

\begin{table}
\centering
\begin{tabular}{l c l c c c c c c c c c c}
\toprule
 & $\mathrm{SU(5)}$ & $\mathrm{A_5}$ & $\zz_{4}^R$ & $\zz_2$ & $\zz_2$ &
$\zz_3$ & $\zz_3$ & $\zz_3$ & $\zz_3$ & $\zz_3$ & $\zz_3$ & $\zz_4$ \\
\midrule
$F                                  $ & $\mathbf{\bar{5}} $ & $\mathbf{3}$ & $1$ & $0$ & $0$ & $0$ & $0$ & $1$ & $2$ & $0$ & $0$ & $0$ \\
$N                                  $ & $\mathbf{1}       $ & $\mathbf{3}$ & $1$ & $0$ & $0$ & $0$ & $0$ & $0$ & $2$ & $0$ & $0$ & $2$ \\
$T_1                                $ & $\mathbf{10}      $ & $\mathbf{1}$ & $1$ & $1$ & $0$ & $2$ & $2$ & $2$ & $2$ & $0$ & $0$ & $0$ \\
$T_2                                $ & $\mathbf{10}      $ & $\mathbf{1}$ & $1$ & $0$ & $0$ & $0$ & $2$ & $1$ & $1$ & $0$ & $0$ & $3$ \\
$T_3                                $ & $\mathbf{10}      $ & $\mathbf{1}$ & $1$ & $0$ & $0$ & $0$ & $0$ & $2$ & $2$ & $0$ & $0$ & $3$ \\
$H_5                                $ & $\mathbf{5}       $ & $\mathbf{1}$ & $0$ & $0$ & $0$ & $0$ & $0$ & $2$ & $2$ & $0$ & $0$ & $2$ \\
$\bar{H}_5                          $ & $\mathbf{\bar{5}} $ & $\mathbf{1}$ & $0$ & $0$ & $0$ & $2$ & $1$ & $2$ & $0$ & $0$ & $1$ & $0$ \\
\midrule
$\phi_3                             $ & $\mathbf{1}$ & $\mathbf{3 }$ & $0$ & $1$ & $1$ & $0$ & $2$ & $0$ & $2$ & $2$ & $0$ & $1$ \\
$\omega                             $ & $\mathbf{1}$ & $\mathbf{5 }$ & $0$ & $0$ & $0$ & $0$ & $0$ & $0$ & $2$ & $0$ & $0$ & $0$ \\
$\theta_2                           $ & $\mathbf{1}$ & $\mathbf{1 }$ & $0$ & $1$ & $1$ & $0$ & $2$ & $1$ & $2$ & $1$ & $0$ & $3$ \\
$\epsilon_1                         $ & $\mathbf{1}$ & $\mathbf{1 }$ & $0$ & $0$ & $0$ & $0$ & $1$ & $1$ & $1$ & $0$ & $0$ & $0$ \\
\midrule
$\Gamma_1                           $ & $\mathbf{1}       $ & $\mathbf{3}$ & $0$ & $0$ & $0$ & $0$ & $1$ & $1$ & $1$ & $0$ & $0$ & $0$ \\
$\bar{\Gamma}_1                     $ & $\mathbf{1}       $ & $\mathbf{3}$ & $2$ & $0$ & $0$ & $0$ & $2$ & $2$ & $2$ & $0$ & $0$ & $0$ \\
$\Gamma_3                           $ & $\mathbf{1}       $ & $\mathbf{1}$ & $0$ & $0$ & $0$ & $0$ & $2$ & $2$ & $2$ & $0$ & $0$ & $0$ \\
$\bar{\Gamma}_3                     $ & $\mathbf{1}       $ & $\mathbf{1}$ & $2$ & $0$ & $0$ & $0$ & $1$ & $1$ & $1$ & $0$ & $0$ & $0$ \\
$\Psi_1                           $ & $\mathbf{5}      $ & $\mathbf{3}$ & $1$ & $0$ & $0$ & $0$ & $1$ & $0$ & $2$ & $0$ & $0$ & $0$ \\
$\bar{\Psi}_1                     $ & $\mathbf{\bar{5}}$ & $\mathbf{3}$ & $1$ & $0$ & $0$ & $0$ & $2$ & $0$ & $1$ & $0$ & $0$ & $0$ \\
$\Psi_2                           $ & $\mathbf{5}      $ & $\mathbf{3}$ & $0$ & $0$ & $0$ & $0$ & $1$ & $0$ & $0$ & $0$ & $0$ & $2$ \\
$\bar{\Psi}_2                     $ & $\mathbf{\bar{5}}$ & $\mathbf{3}$ & $2$ & $0$ & $0$ & $0$ & $2$ & $0$ & $0$ & $0$ & $0$ & $2$ \\
\bottomrule
\end{tabular}
\caption{Charges under $\mathbb{Z}_n$ and $\mathrm{SU(5)}$ and
  $\mathrm{A_5}$ representations of all the fields appearing in
  the neutrino sector of the model. Note that the only new fields
  compared to \cite{Gehrlein:2014wda} are the messenger fields
  $\Psi_1$, $\bar{\Psi}_1$, $\Psi_2$ and $\bar{\Psi}_2$.
  }
\label{tab:Fields}
\end{table}

We are not going to discuss here the quark sector, 
it was analysed in  \cite{Gehrlein:2014wda}. 
In what concerns the charged lepton sector, we only note that 
for the matrix of charged lepton Yukawa couplings we find
\begin{align}
Y_e &=  \begin{pmatrix}
0 & - 1/2 a_{21} & 0\\
6 a_{12} & 6 a_{22} & 6 a_{32}\\
0 & 0 & - 3/2 a_{33}\\
\end{pmatrix}~\; .
\label{eq:yuk_charged}
\end{align}
The order one coefficients in front of the parameters
$a_{ij}$ are SU(5) Clebsch-Gordan coefficients which imply
that
\begin{equation}
\theta_{13} \approx \frac{1}{\sqrt{2}} \theta_{C}  \;,
\label{eq:theta_C1}
\end{equation}
where $\theta_C$ is the Cabibbo angle. This is only
possible due to the non-standard Clebsch-Gordan coeffcients
\cite{Antusch:2009gu} as it was realised
in a series of papers \cite{Antusch:2011qg}.

The flavon $\omega$ is responsible for the GR structure of the Majorana
mass matrix which can be seen in the LO superpotential for the neutrino
sector which reads
\begin{align}
\mathcal{W}^{\text{LO}}_\nu &= y_1^n F N H_5 + y_2^n N N \omega \;.
\label{eq:superpot_neutrino}
\end{align}
The right-handed neutrino mass matrix then reads
\begin{align}
 M_{\text{RR}} &= y_2^n \begin{pmatrix}
2 \sqrt{\frac{2}{3}}(v_2 + v_3) & -\sqrt{3} v_2 & -\sqrt{3} v_2 \\
-\sqrt{3}v_2 & \sqrt{6}v_3 &- \sqrt{\frac{2}{3}}(v_2 + v_3)\\
-\sqrt{3}v_2 & -\sqrt{\frac{2}{3}}(v_2 + v_3) & \sqrt{6} v_3\\
\end{pmatrix} \;
\label{eq:massmatrix_righthanded}
\end{align}
and the neutrino Yukawa couplings are
\begin{equation}
 Y_\nu^{\text{LO}} = y_1^n \begin{pmatrix} 1 & 0 & 0 \\ 0 & 0 & 1 \\ 0 & 1 & 0  \end{pmatrix} \;,
 \label{eq:yukawamatrix_neutrino}
\end{equation}
which are diagonalised by the golden ratio mixing matrix
$U_{\text{GR}}$ from eq.~\eqref{eq:U_GR}.
Note that we are using the right-left convention for the Yukawa matrices,
which means that the first index of the matrix corresponds to the
SU(2)$_L$ singlet.

This is the structure of the original model which 
cannot accommodate for the observed value of the baryon asymmetry of
the universe, as we will see later on.
In order to generate a
non-zero asymmetry we follow the approach as described in \cite{Hagedorn:2009jy}
and introduce an additional operator which perturbs the original flavour structure
of the model. Note nevertheless, that compared to \cite{Hagedorn:2009jy}
we do not introduce an additional flavon and no additional shaping symmetries.
Here it is sufficient to extend minimally
the messenger content of the model.

\subsection{The neutrino sector at NLO}

In this section we discuss the NLO superpotential of the neutrino sector. In order to
accommodate leptogenesis
which can generate the observed value of the baryon asymmetry,
we will introduce a correction to the neutrino Yukawa matrix
governed by the operator $F N \phi_3 \th_2 H_5\epsilon_1^2$. This operator was absent
in the original model but can be added by introducing only two new pairs of
messenger fields $\Psi_1$, $\bar{\Psi}_1$, $\Psi_2$ and $\bar{\Psi}_2$. 

\begin{figure}
\centering
\includegraphics[scale = 0.7]{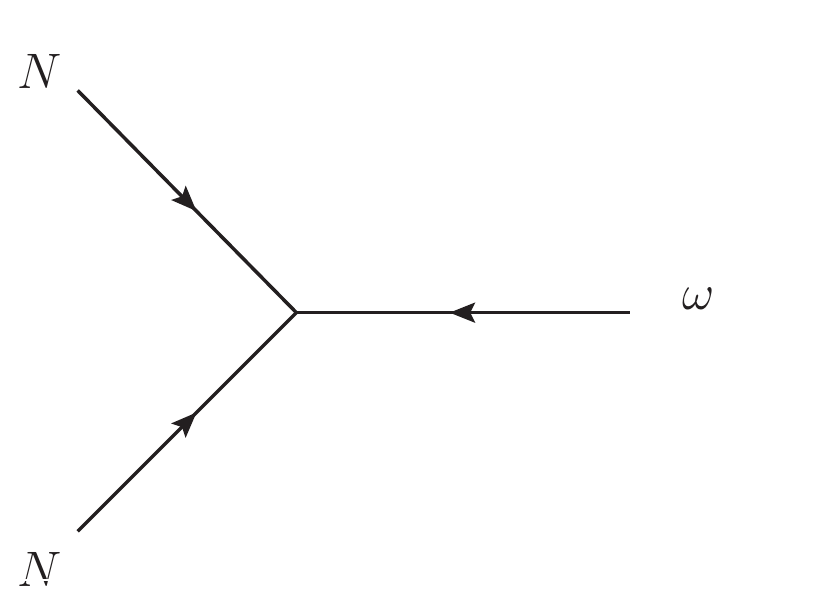}
\includegraphics[scale = 0.7]{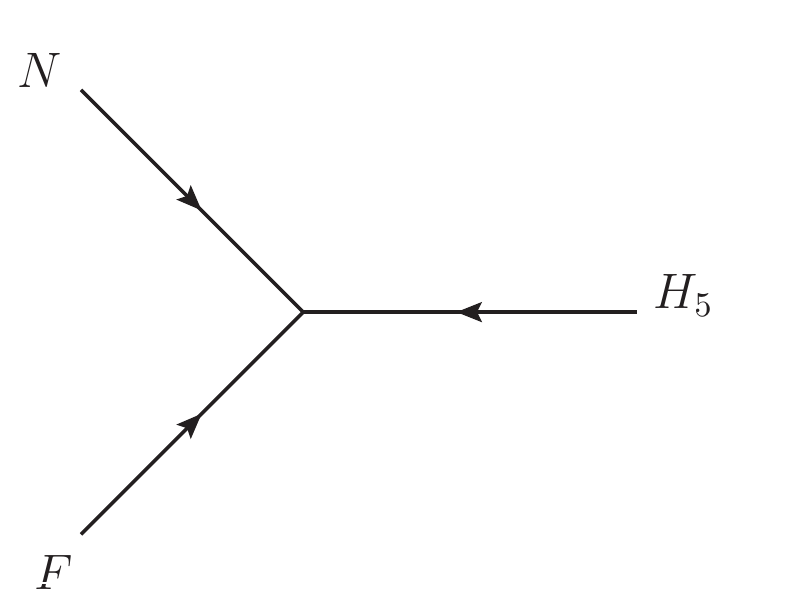} \\[2em]
\includegraphics[scale = 0.7]{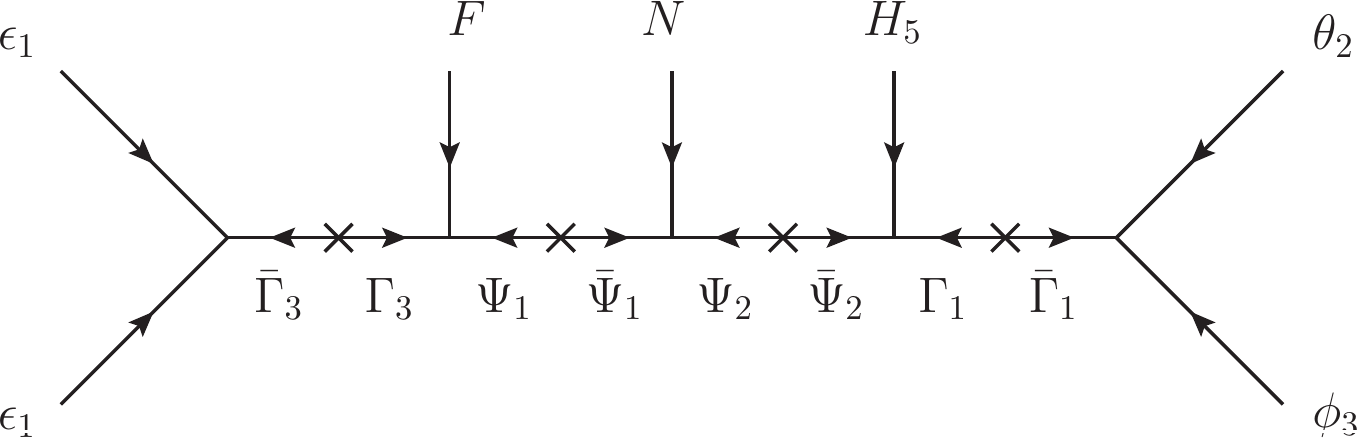}
\caption{
The supergraphs for the neutrino sector including LO and NLO operators.
}
\label{fig:diagram_NLO}
\end{figure}

The renormalisable superpotential of the NLO neutrino sector reads
\begin{align}
\begin{split}
  \mathcal{W}_\nu^{\text{ren, NLO}} &= M_{\Psi_1} \Psi_1 \bar{\Psi}_1 + M_{\Psi_2} \Psi_2 \bar{\Psi}_2 + F \Psi_1 \Gamma_3+N \bar{\Psi}_1 \Psi_2+\bar{\Psi}_2 H_5  \Gamma_1+ \epsilon_1 \epsilon_1 \bar{\Gamma}_3+\theta_2 \phi_3 \bar{\Gamma}_1 \;,
\end{split}
\label{eq:renWnu}
\end{align} 
where we have omitted the coupling constants to increase clarity and only
write down the operators which are new.
We assume the messenger masses to be larger than the GUT scale and to
be related to the messenger scale $\Lambda$ by $\mathcal{O}(1)$ coefficients.
The charges of the new messenger fields under the $\mathbb{Z}_n$ symmetries as well as their SU(5)
and A$_5$ representations are shown in Tab.~\ref{tab:Fields}.
The corresponding supergraphs for the leading order operator and the
next-to-leading order operator for the neutrino Yukawa matrix can be found in
Fig.~\ref{fig:diagram_NLO}.

No new operators  compared
to the original model are possible apart from the one we discuss now. The only new effective operator is
\begin{equation}
\mathcal{W}_\nu^{\text{NLO}} = \frac{1}{\Lambda^4} ( (N F)_{\bf 3} \phi_3)_{\bf 1} H_5 \theta_2 \epsilon_1^2 \; ,
\end{equation}
where we have denoted with brackets the A$_5$ contractions. This operator gives a correction
to the neutrino Yukawa matrix
\begin{equation}
\delta Y_\nu \equiv | y_1^n | c \text{ e}^{\text{i } \gamma} \begin{pmatrix}0&1&0\\-1&0&0\\0&0&0\\ \end{pmatrix} \;,\label{eqn:Noperator}
\end{equation}
where $0 < c \ll 1$. This correction disturbs the golden ratio mixing
pattern already in the neutrino sector by itself and subsequently the
phenomenology of the original model, especially the prediction for
leptogenesis is modified.

\section{Phenomenology: Analytical Results}

In this section we will discuss the phenomenological implications of
introducing $\delta Y_\nu$. Because $c$ is small, in many cases the results
are similar to those obtained in the original model. However, as we will see, 
in some cases when the leading order result was relatively small, 
a correction of order $c$ can have a sizeable impact.

\subsection{Masses and Mixing Angles}

The original model was very predictive due to the built-in sum rules. And indeed
one mass sum rule remains valid. Since $M_{\text{RR}}$ is not corrected
the sum rule for the right-handed neutrino masses
\begin{equation}
M_1 + M_2 = M_3 \;,
\end{equation} 
is still correct. Note that, the masses are taken here to be complex.

The situation for the light 
neutrino masses is somewhat different: 
they get corrections of the order of $c^2 \ll 1$. However, 
since these corrections are small,
the sum rule
\begin{equation}
\frac{\text{e}^{\text{i} \, \alpha_{1}}}{\left|m_{1}\right|}+\frac{\text{e}^{\text{i} \, \alpha_{2}}}{\left|m_{2}\right|} - \frac{1}{\left|m_{3}\right|} = \mathcal{O}(c^2) \approx 0 
\label{eq:masses_sum}
\end{equation}
is still a good approximation. And hence our estimate for the 
ranges of the neutrino masses from \cite{Gehrlein:2014wda}
\begin{align}
  0.011 \text{ eV} &\lesssim  m_1 \phantom{\lesssim 0.454 \text{ eV}} \text{  for NO,} \label{eq:NHMassRange} \\
  0.028 \text{ eV} &\lesssim  m_3 \lesssim 0.454 \text{ eV} \text{ for IO.} \label{eq:IHMassRange}
\end{align}
is still reasonable.
For all three PMNS mixing angles we find corrections of order $c$.
For $\theta_{13}$ and $\theta_{23}$ the expressions are somewhat
lengthy and not insightful, but as an example we find as correction
for $\theta_{12}$ to first order in $c$

\begin{align}
\delta \theta_{12}^\nu &= - \frac{c \cos \gamma}{\sqrt{2}} - \frac{\sqrt{2} c  \Im(M_1^\star M_2) \sin \gamma }{|M_1^2| - |M_2^2|} \; .
\end{align}
The corrections have immediate consequences for the phenomenology.

The first thing one might wonder, is if 
the inverted mass ordering is still excluded like in the original model.
We begin our discussion
with the sum rule \cite{Petcov:2014laa}\footnote{
In the original model we had used another sum rule from \cite{Antusch:2005kw}
which can be derived from this sum rule by expanding in $\theta_{13}$.
But since $\theta_{13}$ is not very small we want to use now the improved
sum rule.}
\begin{equation}
\sin^2\theta_{12} = \cos^2\theta^{\nu}_{12} +
\frac{\sin2\theta_{12}\sin\theta_{13}\cos\delta
 -  \tan\theta_{23}\cos2\theta^{\nu}_{12}}
{\tan\theta_{23} (1 - \cot^2\theta_{23}\,\sin^2\theta_{13})}\, ,
\label{s2th12cosdthnu}
\end{equation}
where $\cos^2\theta^{\nu}_{12} = \phi_{g}/\sqrt{5}$ and $\cos2\theta^{\nu}_{12} = 2\phi_{g}/\sqrt{5} - 1$.

For $c = 0$ we can evaluate this sum rule and find $\theta_{12} \gtrsim 23^\circ$
(compared to $\theta_{12} \gtrsim 24^\circ$ from the sum rule in \cite{Antusch:2005kw}).
Using the lower bound on the mass scale in the inverted ordering case, cf.\ eq.~\eqref{eq:IHMassRange},
we can estimate that
the RGE evolved value of $\theta_{12}$ at the seesaw scale has to be smaller than about
5.7$^\circ$. Attributing this difference completely to the correction $\delta \theta_{12}^\nu$
with $\gamma = 0$ or $\pi$
we find $c \gtrsim 0.43$.
This is a crude estimate because the other mixing angles are affected as well
modifying the above sum rule in a non-trivial way. Nevertheless, from here
we would still expect
$c$ to be of order $0.1$ to safe the inverted mass ordering. On the other hand such a large
value for $c$ is not plausible from a model building point of view because it is associated
to a highly suppressed operator making values of $c = 10^{-4}$ to $10^{-3}$ plausible.
We will come back to this point later when we discuss the numerical results.

For future convenience, we introduce the following definitions for the
parameters which we will use as well in our numerical scan
\begin{align}
M_1&=\frac{1}{\sqrt{6}}(X+Y)=\frac{1}{\sqrt{6}}|X||1+\rho \text{ e}^{\text{i}\phi}|\text{e}^{\text{i}\phi_1}, ~\phi_1=\text{arg}(X+Y)~,\\
M_2&=\frac{1}{\sqrt{6}}(X-Y)=\frac{1}{\sqrt{6}}|X||1-\rho \text{ e}^{\text{i}\phi}|\text{e}^{\text{i}\phi_2},~ \phi_2=\text{arg}(X-Y)~,\\
M_3&=\sqrt{\frac{2}{3}} X=\sqrt{\frac{2}{3}} |X|\text{ e}^{\text{i}\phi_3},~ \phi_3=\text{arg}(X)~,
\end{align}
where 
\begin{align}
X&=(4 v_3+ v_2)y_2^n~,\\
Y&=3\sqrt{5}v_2 y_2^n~,\\
\rho&=\left|\frac{Y}{X}\right|~,\\
\phi&=\text{arg}(Y)-\text{arg}(X)~.
\end{align}

In this way, we express the absolute value of the three heavy neutrino masses in terms of three real parameters, i.e., $|X|$, $\rho$ and $\phi$. $|X|$ sets the scale of our interest. $\rho$ reflects the detailed structure of the heavy neutrino mass spectrum. $\phi$ is connected to $\rho$ via the ratio of two mass squared differences
\begin{eqnarray}
\frac{\Delta m_{21}^2}{\Delta m_{31}^2}=\frac{16 \rho \cos\phi}{(\rho^2-2\rho\cos\phi+1)(\rho^2+2\rho\cos\phi-3)}~.\label{alphaphi}
\end{eqnarray}
Notice that we neglect here for the moment RG effects on the masses
and corrections of order $c^2$ to the neutrino masses which will turn
out to be well justified in the numerical analysis.

\begin{figure}
\centering
\includegraphics[scale=0.5]{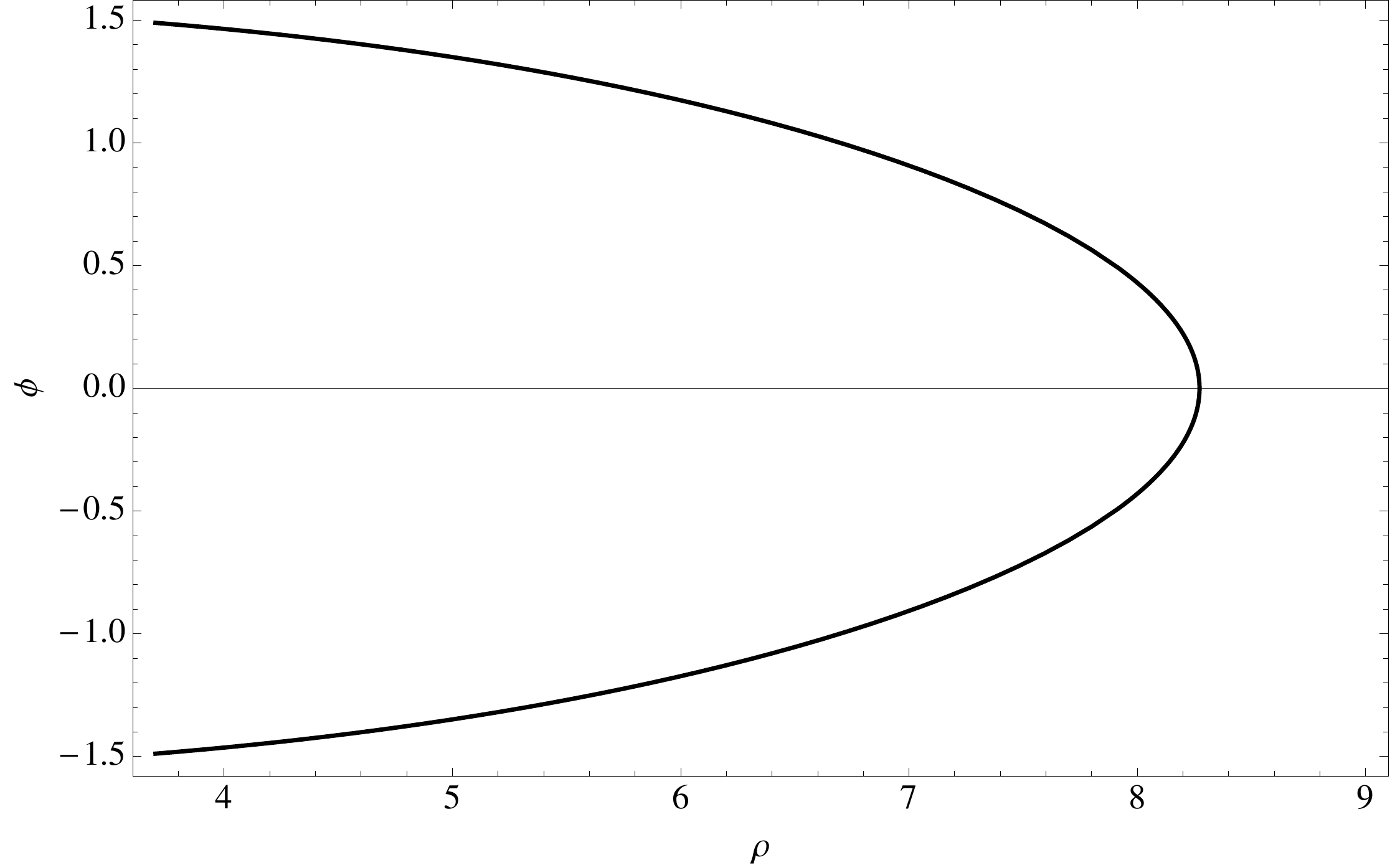} \\[1em]
\includegraphics[scale=0.69]{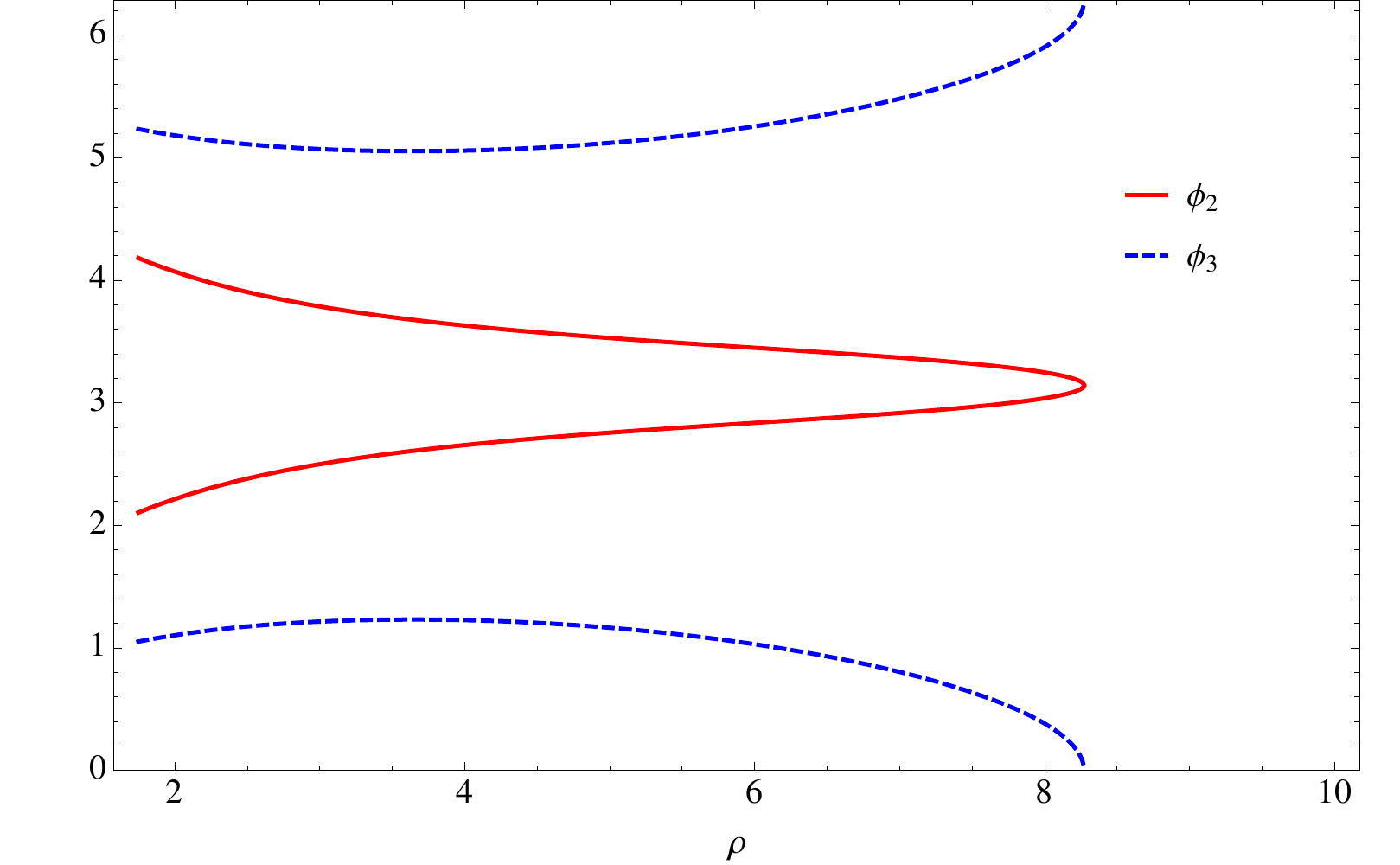}
\caption{
  The dependence of $\phi$ and the phases of the heavy Majorana
  neutrinos on $\rho$ according to eqs.~(\ref{alphaphi}, \ref{alphaphi2}, \ref{alphaphi3}).
 The unit for $\phi, \phi_2, \phi_3$ is rad. We use the best fit values from the global
 fit \cite{Gonzalez-Garcia:2014bfa} for the two squared mass differences as input here.
}
\label{fig:PhiAlpha}
\end{figure}

One of the Majorana phases which we choose to be $\phi_1$ can be set to zero by
applying a redefinition of the heavy Majorana fields.
The remaining two phases $\phi_2$ and $\phi_3$ can as well be expressed in terms of
$\rho$ and $\phi$ using the complex mass sum rule $M_1+M_2=M_3$
\begin{align}
\cos\phi_2&=\frac{|M_3|^2-|M_1|^2-|M_2|^2}{2|M_1||M_2|}=\frac{1-\rho^2}{\sqrt{1-2\rho^2\cos 2\phi+\rho^4}}~,\label{alphaphi2}\\
\cos\phi_3&=\frac{|M_1|^2-|M_2|^2+|M_3|^2}{2|M_1||M_3|}=\frac{1+\rho \cos\phi}{\sqrt{1+2\rho\cos \phi+\rho^2}}~.\label{alphaphi3}
\end{align}
Notice that only normal ordering is viable in this model, and the Yukawa
couplings are degenerate in LO so that we have $|M_3|<|M_2|<|M_1|$.
Thus $\cos\phi$ is positive, $\cos\phi_2$ is negative, $\cos\phi_3$ is positive, which
gives us a first constraint on $\rho$ which we will comment on later. Notice also that
the sign of $\sin\phi_2$ and $\sin\phi_3$ is not fixed. We plot the dependence of the
phases on $\rho$ in Fig.~\ref{fig:PhiAlpha}.
The Majorana phases $\alpha_1$, $\alpha_2$
and $\phi_2$, $\phi_3$ are related via
\begin{align} \label{eq:MajoranaPhases}
 \alpha_1 = -\phi_3 \text{ and } \alpha_2 = \phi_2 - \phi_3 \;.
\end{align}
up to order $c^2$.

We comment a little on the phases in the mass matrices. The heavy neutrino mass matrix is diagonalised as 
\begin{align}
U_{ \text{GR}}^{\text{T}} M_{\text{RR}} U_{ \text{GR}}= D_N=\rm \text{Diag}( M_1\text{ e}^{\text{i}\phi_1},M_2\text{ e}^{\text{i}\phi_2},M_3\text{ e}^{\text{i}\phi_3} )\;,~\quad M_{1,2,3}>0~.
\end{align}
We eliminate the common phase by setting $\phi_1=0$ and attribute the phase factors to a phase matrix $ P=\text{Diag}(1, \text{e}^{\text{i}\phi_2/2}, \text{e}^{\text{i}\phi_3/2} ) $. Thus we have
\begin{align}
P^{-1} U_{ \text{GR}}^{\text{T}} M_{\text{RR}} U_{\text{GR}} P^{-1}= \rm \text{Diag}(M_1,M_2,M_3) \;,
\end{align} 
which means $U_{\text{GR}}P^{-1}$ diagonalises the heavy neutrino mass matrix to real and non-negative eigenvalues. Applying the seesaw mechanism, we have
\begin{align}
m_{\text{LL}}=-m_{\text{D}}^{\text{T}} M_{\text{RR}}^{-1} m_{\text{D}}=U_\nu^\star \text{Diag} (m_1,m_2,m_3) U_\nu^\dagger~.\label{lightnumass}
\end{align}
Notice that from $U_{ \text{GR}}^{\text{T}} m_{\text{D}} U_{\text{ GR}}= y_1^n v_u\rm \text{Diag} (1,1,-1) + \mathcal{O}(c)$ we get
\begin{align}
U_\nu= \text{i } U_{\text{GR}} P~ + \mathcal{O}(c).
\end{align}
If we choose $y_1^n$ to be real and positive, the only complexity comes from the heavy
neutrino mass matrix and the phase $\gamma$. By ascribing the phases to the $P$ matrix,
the $m_i$, $i=1,2,3$ in eq.~\eqref{lightnumass} are real and non-negative up to
corrections of $\mathcal{O}(c)$. From now on, we use the symbol $M_i$ and $m_i$ to label
the real and non-negative masses.

\subsection{Leptogenesis}
\label{sec:AnalyticalLeptogenesis}

In this section we discuss analytical estimates for the generated baryon asymmetry
including all relevant parameters and formulas. 
To discuss leptogenesis in this model, we first set our scale of interest, the see-saw
scale, to be $M_S\simeq10^{13}$ GeV. Taking into consideration $\tan\beta=30$, we
have $10^9 (1+\tan^2\beta)$ GeV$<M_S<10^{12} (1+\tan^2\beta)$~GeV, which
as was shown in \cite{Pascoli:2006ci},
corresponds to the ``two-flavoured leptogenesis''
regime~\cite{Nardi:2006fx,Abada:2006fw,Abada:2006ea}, i.e., the
regime where the 
processes mediated by the $\tau$ Yukawa 
couplings enter into equilibrium.
Later on in our numerical scan we will find 
that in order to generate realistic neutrino masses
the parameter $\rho$ has to satisfy the inequality
$\rho \gtrsim 5.8$. This in turn implies that 
the leptogenesis regime in the model we are considering 
cannot be resonant. Indeed, as can be shown, 
for $\rho\gtrsim 3.7$ we have 
$M_1-M_2\gg \Gamma_1= 
\tfrac{(\tilde{Y}_\nu \tilde{Y}_\nu^\dagger)_{11} M_1}{8\pi}$, 
the smallest heavy Majorana neutrino mass splitting $(M_1 - M_2)$
being by at least two orders of magnitude larger 
than  $\Gamma_1$. Thus,
the condition of resonant leptogenesis \cite{Pilaftsis:1997jf}
$(M_1-M_2)\sim \Gamma_1$, is not satisfied in the model under discussion.

The CP-asymmetry generated in the lepton charge $L_{l}$ 
by neutrino and sneutrino decays, $l = e,\mu,\tau$, 
is~\cite{Davidson:2008bu}:
\begin{align}
\epsilon_i^{l}=&\frac{1}{8\pi}\frac{1}{(\tilde{Y}_\nu \tilde{Y}_\nu^\dagger)_{ii}}\sum_{j\neq i} {\rm Im}[ (\tilde{Y}_\nu)_{j l}(\tilde{Y}_\nu)_{i l }^\star (\tilde{Y}_\nu \tilde{Y}_\nu^\dagger)_{ji}] f ( \tfrac{m_i}{m_j} )\nonumber\\
&+\frac{1}{8\pi}\frac{1}{(\tilde{Y}_\nu \tilde{Y}_\nu^\dagger)_{ii}}\sum_{j\neq i} {\rm Im }[ (\tilde{Y}_\nu)_{jl }(\tilde{Y}_\nu)_{i l}^\star (\tilde{Y}_\nu \tilde{Y}_\nu^\dagger)_{ij} ] \frac{m_j^2}{m_j^2-m_i^2}~,\label{epsilon}
\end{align}
where 
\begin{align}
f(x)=-x \left( \frac{2}{x^2-1}+\text{log}(1+\frac{1}{x^2}) \right)~.
\end{align}

The second term in eq.~\eqref{epsilon} corresponds to the self-energy diagram 
with an inverted fermion line in the loop. It would vanish when we sum over $\alpha$,
and we would end up with the same formula as in the one flavour case
\begin{align}
 \epsilon_i=\sum_l \epsilon_i^l =\frac{1}{8\pi}\sum_{j\neq i}\frac{ {\rm Im}[(\tilde{Y}_\nu\tilde{Y}_\nu^\dagger)_{ji}^2]}{(\tilde{Y}_\nu\tilde{Y}_\nu^\dagger)_{ii}} f(\tfrac{m_i}{m_j})~.
\end{align}

In the basis where the charged lepton and the right handed neutrino mass matrices are diagonal, we have
\begin{align}
\tilde{Y}_\nu&= (U_{\text{GR}}P^{-1})^\dagger (Y_\nu+\delta Y_\nu)U_e\nonumber\\
&= y_1^n\left(
\begin{array}{ccc}
 -\frac{\left(s^e_{12} \text{e}^{\text{i} \delta_{12}^e}+\sqrt{3+\sqrt{5}} c^e_{12}\right) }{\sqrt{5+\sqrt{5}}} & \frac{\text{e}^{-\text{i} \delta_{12}^e} \left(c^e_{12} \text{e}^{\text{i} \delta_{12}^e}-\sqrt{3+\sqrt{5}} s^e_{12}\right) }{\sqrt{5+\sqrt{5}}} & \frac{1}{\sqrt{5+\sqrt{5}}} \\
 \frac{\text{e}^{\frac{\text{i} \phi_2}{2}} \left(\left(5+\sqrt{5}\right) \text{e}^{\text{i} \delta_{12}^e} s^e_{12}-2 \sqrt{10} c^e_{12}\right) }{2 \sqrt{5 \left(5+\sqrt{5}\right)}} & -\frac{\text{e}^{\frac{\text{i} \phi_2}{2}-\text{i} \delta_{12}^e} \left(\left(5+\sqrt{5}\right) c^e_{12} \text{e}^{\text{i}\delta_{12}^e}+2 \sqrt{10} s^e_{12}\right) }{2\ 5^{3/4} \sqrt{1+\sqrt{5}}} & -\frac{1}{2} \sqrt{1+\frac{1}{\sqrt{5}}} \text{e}^{\frac{\text{i} \phi_2}{2}}  \\
 -\frac{\text{e}^{\frac{1}{2} \text{i} (2 \delta_{12}^e+\phi_3)} s^e_{12} }{\sqrt{2}} & \frac{c^e_{12} \text{e}^{\frac{\text{i} \phi_3}{2}} }{\sqrt{2}} & -\frac{\text{e}^{\frac{\text{i} \phi_3}{2}} }{\sqrt{2}} \\
\end{array}
\right)\nonumber\\
&+ y_1^n c \, \text{e}^{\text{i} \gamma}\left(
\begin{array}{ccc}
 -\frac{ \left(c^e_{12}-\sqrt{3+\sqrt{5}} \text{e}^{\text{i} \delta_{12}^e} s^e_{12}\right)}{\sqrt{5+\sqrt{5}}} & -\frac{ \text{e}^{-\text{i} \delta_{12}^e} \left(\sqrt{3+\sqrt{5}} c^e_{12} \text{e}^{\text{i} \delta_{12}^e}+s^e_{12}\right)}{\sqrt{5+\sqrt{5}}} & 0 \\
 \frac{ \text{e}^{\frac{\text{i} \phi_2}{2}} \left(2 \sqrt{2} s^e_{12} \text{e}^{\text{i} \delta_{12}^e}+\sqrt{5} c^e_{12}+c^e_{12}\right)}{2 \sqrt{5+\sqrt{5}}} & \frac{ \text{e}^{\frac{\text{i} \phi_2}{2}-\text{i} \delta_{12}^e} \left(-2 \sqrt{2} c^e_{12} \text{e}^{\text{i} \delta_{12}^e}+\sqrt{5} s^e_{12}+s^e_{12}\right)}{2 \sqrt{5+\sqrt{5}}} & 0 \\
 \frac{c^e_{12}  \text{e}^{\frac{\text{i} \phi_3}{2}}}{\sqrt{2}} & \frac{ \text{e}^{\frac{\text{i} \phi_3}{2}-\text{i} \delta_{12}^e} s^e_{12}}{\sqrt{2}} & 0 \\
\end{array}
\right)~,\label{y}
\end{align}
where we use $U_e\simeq U_{12}$, and the abbreviations $\sin\theta_{12}^e=s^e_{12}$
and $\cos\theta_{12}^e=c^e_{12}$. 

Here and in the following we have used the freedom to 
redefine $Y_\nu$ by a global phase
to make $y_1^n > 0$ so that we find
\begin{align}
\tilde{Y}_\nu\tilde{Y}_\nu^\dagger&= P U_{\text{GR}}^{\text{T}}(Y_\nu Y_\nu^\dagger+Y_\nu \delta Y_\nu^\dagger+\delta Y_\nu Y_\nu^\dagger)U_{\text{GR}} P^{-1}\\ \notag
&= \left({y_1^n}\right)^2 \mathds{1}
+c \left({y_1^n}\right)^2
   \left(\begin{smallmatrix}
   0 & \text{i} \sqrt{2}\sin\gamma \text{e}^{-\text{i}\phi_2/2} &  -\sqrt{1+\frac{1}{\sqrt{5}}}\cos\gamma \text{e}^{-\text{i}\phi_3/2}\\
   -\text{i} \sqrt{2}\sin\gamma \text{e}^{\text{i}\phi_2/2} & 0  &  -\frac{2}{\sqrt{5+\sqrt{5}}}\cos\gamma \text{e}^{-\text{i}(\phi_3-\phi_2)/2}\\
    -\sqrt{1+\frac{1}{\sqrt{5}}}\cos\gamma \text{e}^{\text{i}\phi_3/2}&  -\frac{2}{\sqrt{5+\sqrt{5}}}\cos\gamma \text{e}^{\text{i}(\phi_3-\phi_2)/2}& 0\\
   \end{smallmatrix}\right)~,\label{yy}
\end{align}
which we have expanded up to $\mathcal{O}(c)$.

We give next the expressions for the 
CP-violating asymmetries  
in the $l$ lepton charge $L_{l}$, generated in the decays 
of the heavy Majorana neutrinos $N_1$, $N_2$ and $N_3$,
as calculated from eq.~\eqref{epsilon}:
\begin{align}
\epsilon_1^\tau&=\frac{c \left({y_1^n}\right)^2}{8 \pi}\frac{1}{\sqrt{10}} \left( \sin\gamma \cos{\phi_2} f(\tfrac{m_1}{m_2})-\sin\gamma \frac{m_2^2}{m_2^2-m_1^2}
+\cos\gamma \sin{\phi_3} f(\tfrac{m_1}{m_3}) \right)~, \\
\epsilon_2^\tau&=\frac{c \left({y_1^n}\right)^2}{8 \pi}\frac{1}{\sqrt{10}}\left( -\sin\gamma \cos{\phi_2} f(\tfrac{m_2}{m_1})+ \sin\gamma \frac{m_1^2}{m_1^2-m_2^2}
-\cos\gamma\sin{(\phi_3-\phi_2)} f(\tfrac{m_2}{m_3}) \right)~, \\
\epsilon_3^\tau&=\frac{c \left({y_1^n}\right)^2}{8 \pi}\frac{1}{\sqrt{10}}\cos\gamma \left( -\sin{\phi_3} f(\tfrac{m_3}{m_1})+ \sin{(\phi_3-\phi_2)} f(\tfrac{m_3}{m_2}) \right)~.
\end{align}
We see that to leading order, $\epsilon_i^\tau=0$ and hence
leptogenesis was not viable in the original model.
As $\tilde{Y}_\nu$ in leading order is unitary (except for an overall factor
$\left({y_1^n}\right)^2$), we have $\epsilon_i^2\equiv\epsilon_i^e+\epsilon_i^\mu=-\epsilon_i^\tau$
to leading order.

Notice that we compute the CP asymmetry generated by all 
three heavy (s)neutrino decays
since the heavy neutrino spectrum is not very hierarchical in our case.

At the leptogenesis scale and values of $y^n_1$ of interest, the $\Delta L = 
2$ processes are
negligible.
They would be important for a different setup with maximal
perturbative values of the Yukawa coupling of interest (say, for $y_1^n \cong 1$)
if the leptogenesis scale would be $10^{14}$ GeV (or for masses of the heavy Majorana
neutrinos of the order of $10^{14}$ GeV).
Thus, we can use the following analytic approximation 
for the efficiency factors~\cite{Abada:2006ea},
which accounts for the $\Delta L=1$ interactions and 
the decoherence effects:\\

\begin{align}
\eta(\tilde{m}_{il})\simeq\left(\left(\frac{\tilde{m}_{il}}{8.25\times 10^{-3}\text{ eV}}\right)^{-1}+\left(\frac{0.2\times 10^{-3} \text{ eV}}{\tilde{m}_{il}}\right)^{-1.16}\right)^{-1}~,
\end{align}
where 
\begin{eqnarray}
\tilde{m}_{il}=\frac{v_u^2|(\tilde{Y}_\nu)_{il}|^2 }{M_i}~,
\end{eqnarray}
where we introduce another index $i$, $i = 1,2,3$, to label the 
correspondence to the $i$-th heavy (s)neutrino and  $l=e,\mu,\tau$.
If we only keep leading order term in $\tilde{m}_{il}$, we will 
have $\tilde{m}_{i2}\equiv\tilde{m}_{ie}+\tilde{m}_{i\mu}=\frac{v_u^2{(y_1^n)}^2}{M_i}-\tilde{m}_{i\tau}$.
We list the washout mass parameters as follows
\begin{align}
\tilde{m}_{1\tau}&=\frac{1}{5+\sqrt{5}}\frac{v_u^2{(y_1^n)}^2}{M_1}~,\\
\tilde{m}_{2\tau}&=\frac{1}{4} \left(1+\frac{1}{\sqrt{5}}\right)\frac{v_u^2{(y_1^n)}^2}{M_2}~,\\
\tilde{m}_{3\tau}&=\frac{1}{2}\frac{v_u^2{(y_1^n)}^2}{M_3}~.
\end{align}
We do not include the higher order terms $\mathcal{O}(c)$ 
in the expressions for $\tilde{m}_{il}$ 
because they generate subleading insignificant corrections.
%%%%%%%%%%%%%%%%%%%%%%%%%%%%%%%%%
\begin{figure}
\centering
\includegraphics[scale=0.7]{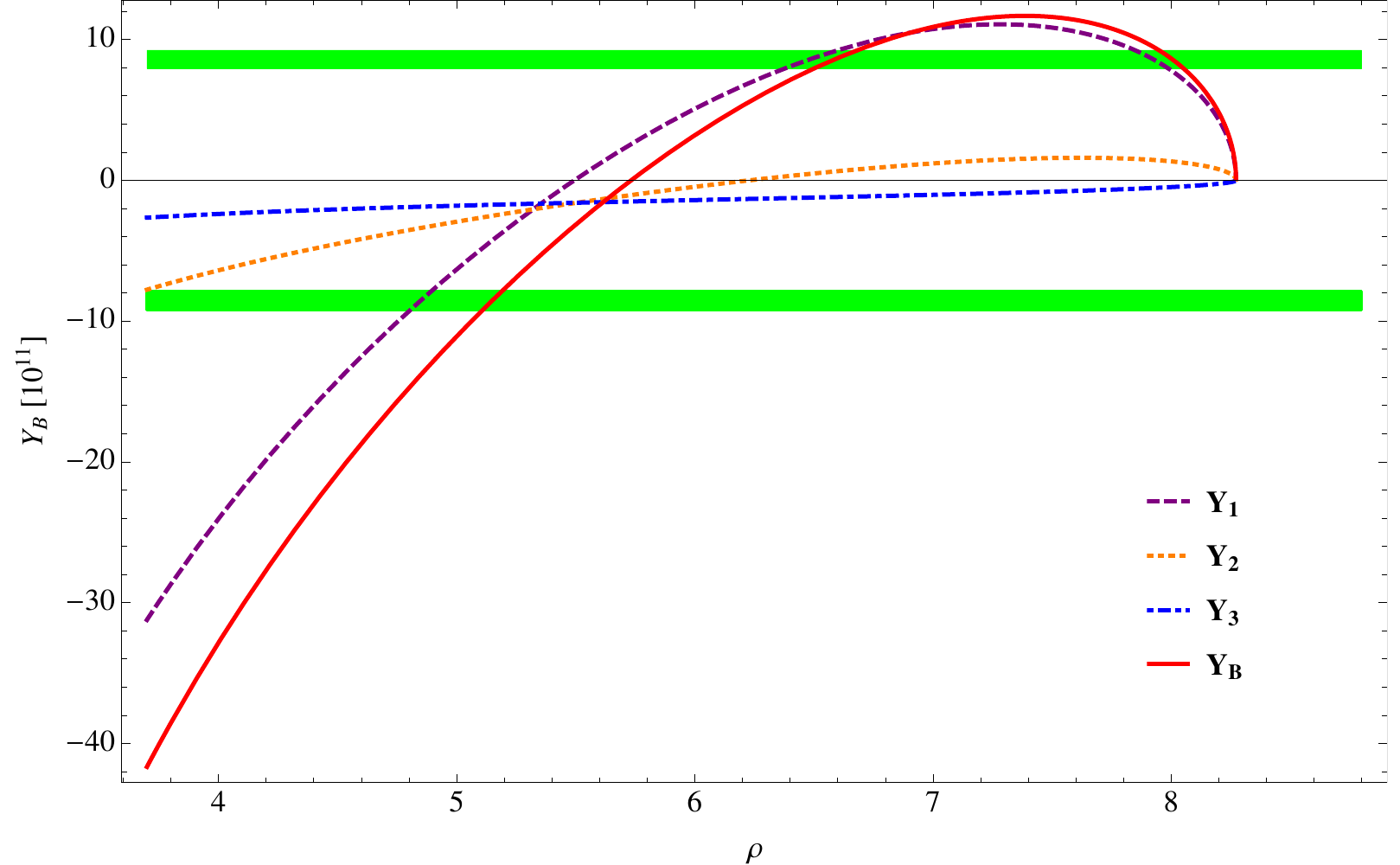}\\
\phantom{a}
\includegraphics[scale=0.7]{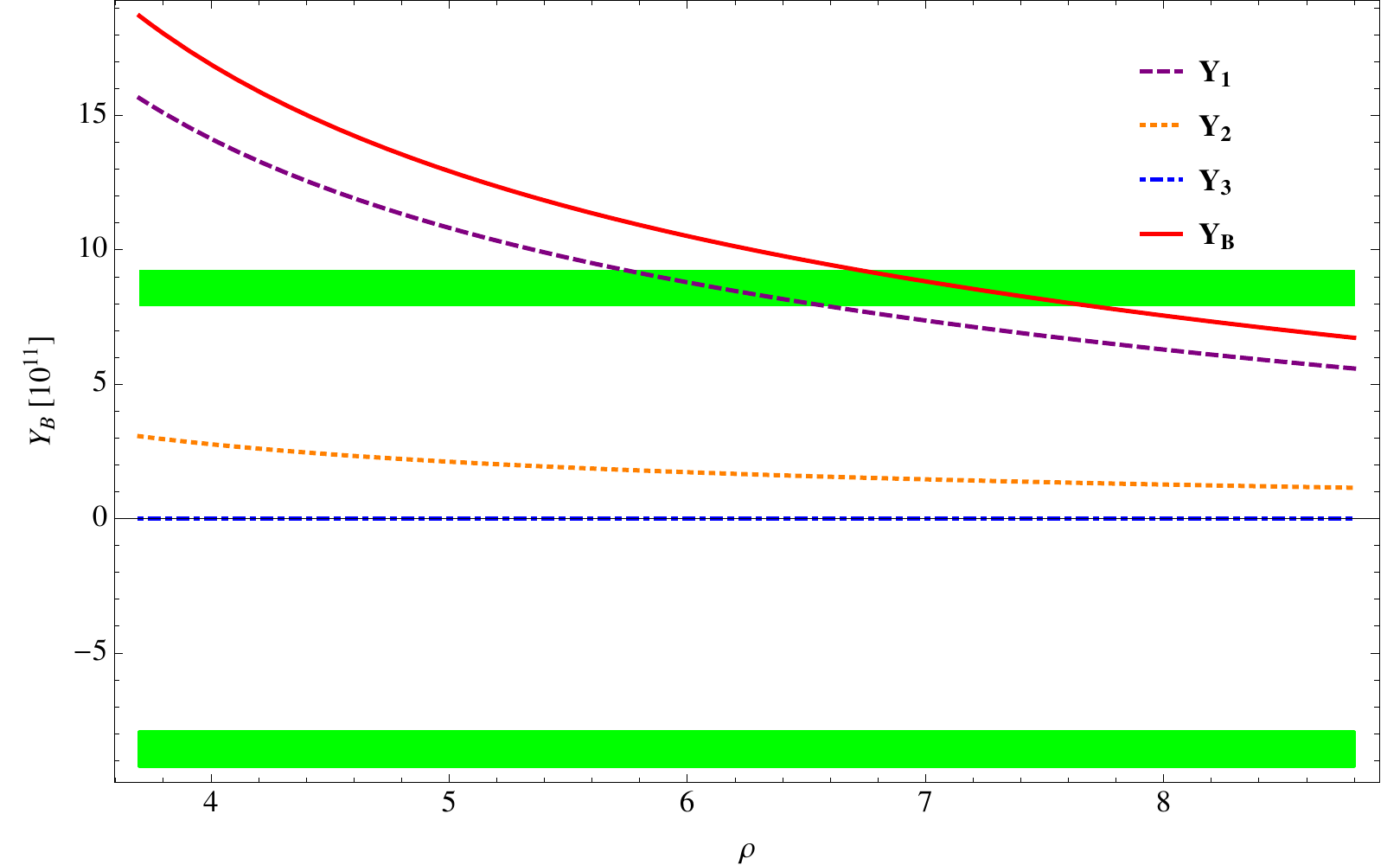}
\caption{
The single asymmetries $Y_1, Y_2, Y_3$ and the total
asymmetry $Y_B$. In the upper plot we use $c=0.05 \approx \theta_C^2$, 
$\gamma=2k\pi ~(k=0,\pm1,\pm2,...)$, $v_u=246$~GeV, $y_1^n=0.1$,
$|X|=10^{13}~\text{GeV}$ and in the lower plot we use
$c=5.8 \cdot 10^{-3} \approx \theta_C^5$, 
$\gamma=\pi/2 + 2k\pi$, $v_u=246$~GeV, $y_1^n=0.1$,
$|X|=7.2 \cdot 10^{12}~\text{GeV}$.
The horizontal green bands correspond
to the 3$\sigma$ region for the observed value for
$|Y_B|=(8.58\pm0.22) \times 10^{-11}$, where we multiply for
the 3$\sigma$ region the 1$\sigma$ error
for the sake of simplicity by a factor of three.
}
\label{fig:y_t}
\end{figure}
%%%%%%%%%%%%%%%%%%%%%%%%%%%%%%%%
The baryon asymmetry generated by each heavy neutrino 
decay is~\cite{Pascoli:2006ci}
\begin{eqnarray}
Y_i\simeq-3\times 10^{-3}\epsilon_i^\tau \left(\eta\left(\frac{494}{761}\tilde{m}_{i\tau }\right)-\eta\left(\frac{541}{761}\tilde{m}_{i2}\right)\right),
\end{eqnarray}
and the total baryon asymmetry is
\begin{eqnarray}
Y_B=\sum_iY_i,~ i=1,2,3.
\end{eqnarray}
Notice that we use an incoherent sum over the asymmetry generated by each
heavy (s)neutrino. 
 This approximation corresponds, in particular, to neglecting 
the wash-out effects due to the lighter heavy Majorana neutrinos 
$N_{2,3}$ in the asymmetry generated by the heaviest Majorana 
neutrino $N_1$. Thus, we effectively assume that the indicated 
wash-out effects cannot reduce drastically the asymmetry 
produced in the $N_1$ decays. Since the masses of 
$N_{2,3}$ and $N_1$ in the model we are considering 
differ at most by a factor of 5, we can expect 
that at least for some ranges of values of the masses 
of $N_1$ and $N_{2,3}$ the wash-out effects under discussion 
will be subdominant, i.e., will lead to a reduction 
of the asymmetry $Y_1$ at most by a factor of 3.
Such a reduction will still allow a generation of $Y_{\rm B}$ 
compatible with the observations. Accounting quantitatively 
for the wash-out effects of interest requires 
solving numerically the system of Boltzmann equations 
describing the evolution of the $N_{1,2,3}$ number 
densities and of the asymmetries $Y_{1,2,3}$ in the 
Early Universe. Performing such a calculation is beyond the scope 
of the present work; it will be done elsewhere.

A priori, we do not know the value of 
$c \, \text{e}^{\text{i}\gamma}$ introduced in
the NLO operator in eq.~\eqref{eqn:Noperator}. 
We will see in the next section that
the low energy observables combined with $Y_B$ will 
give us information on its value.
For now, as interesting cases used for illustration, 
we plot $Y_B$ for some special
values of the parameters in Fig.~\ref{fig:y_t}. 
The first set/upper plot will turn out to be
not realistic but it is still interesting
because here we can see clearly, that the Majorana 
phases of the heavy right handed
neutrinos are the only sources for CP violation and
sufficient to generate $Y_B$ via leptogenesis. 
In this case the sign flip of $Y_1$ and
$Y_2$ is due to the loop functions. The sign of $\sin\phi_i, i=2,3$ 
can be inferred
from the ``right sign" observation of $Y_B$. The second set/lower 
plot is inspired by
the numerical results later on. Leptogenesis is still
successful although there $c$ is chosen much smaller than 
in the first set, since we
receive contribution from the $\sin\gamma$ term, where 
the enhancement from the loop
functions $f(m_1/m_2)$ and $f(m_2/m_1)$ are included. Specifically, we have
$f(m_1/m_2)\simeq-30 f(m_1/m_3)$ for $\rho=7$. In both cases $Y_B$ is dominated
by $Y_1$. The main difference
between $Y_1$ and $Y_2$ is the efficiency factor:
$\eta|_{Y_1}/\eta|_{Y_2}\simeq 5$ for $\rho = 7$. $Y_3$ suffers from
a strong washout in the first case and is zero in the second case due to $\cos\gamma=0$.
The NLO contribution can be regarded as
an expansion in powers of $\theta_C$ in both cases.

\section{Phenomenology: Numerical Results}

In this section we discuss the numerical results of a parameter scan. The analytical
results give a first impression of the general behaviour of all the observables but
since there are several parameters involved which interplay non-trivially we made a
random scan of the parameter space with certain assumptions to prove
that our model can simultaneously fulfill all the constraints. The structure of this
part follows the structure of the previous section.

\subsection{Masses and Mixing Angles}

For our numerical scan we follow closely the method as described in
\cite{Gehrlein:2014wda}. Most importantly for the parameters describing
the quark and charged lepton sector we used the fit results given there.
This implies that we use here $\tan \beta = 30$ and $M_{\text{SUSY}} = 1$~TeV.

In our previous model we had to scan over four real parameters (two moduli
$|X|$ and $|Y|$, two phases $\phi$ and $\delta_{12}^e$) in the neutrino sector.
 In addition to these we have now
scanned as well over the modulus $c$ and the phase $\gamma$.
And now we have included in our scan as additional constraint
\cite{Ade:2013zuv, Bennett:2012zja}
\begin{equation}
Y_B = (8.58 \pm 0.22) \times 10^{-11} \;,
\end{equation}
where we multiply for the 3$\sigma$ region the 1$\sigma$ error
for the sake of simplicity by a factor of three. For the calculation
of $Y_B$ we use the formulas from section~\ref{sec:AnalyticalLeptogenesis}.

\begin{figure}
\centering
\includegraphics[scale=0.49]{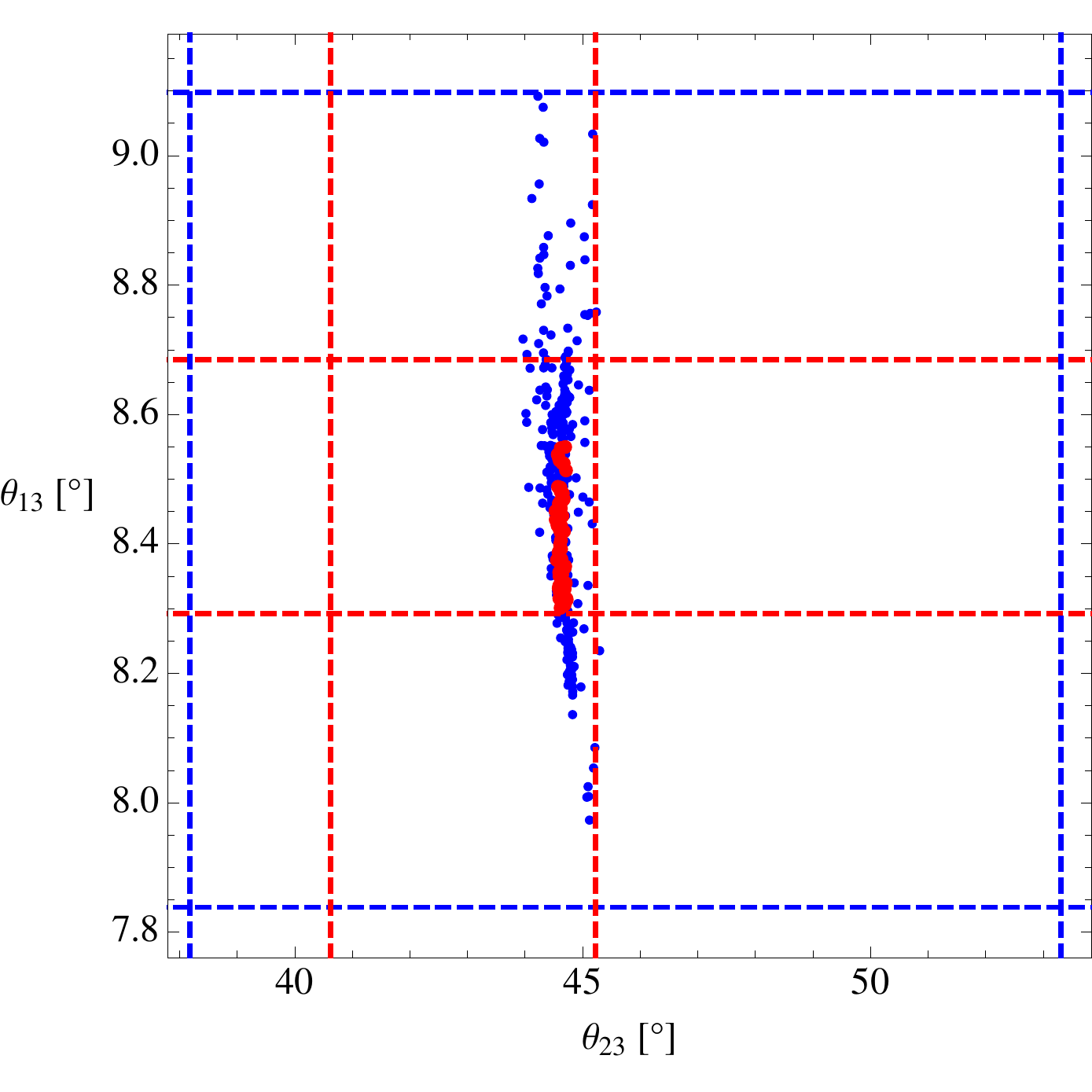} \hspace{0.7cm}
\includegraphics[scale=0.49]{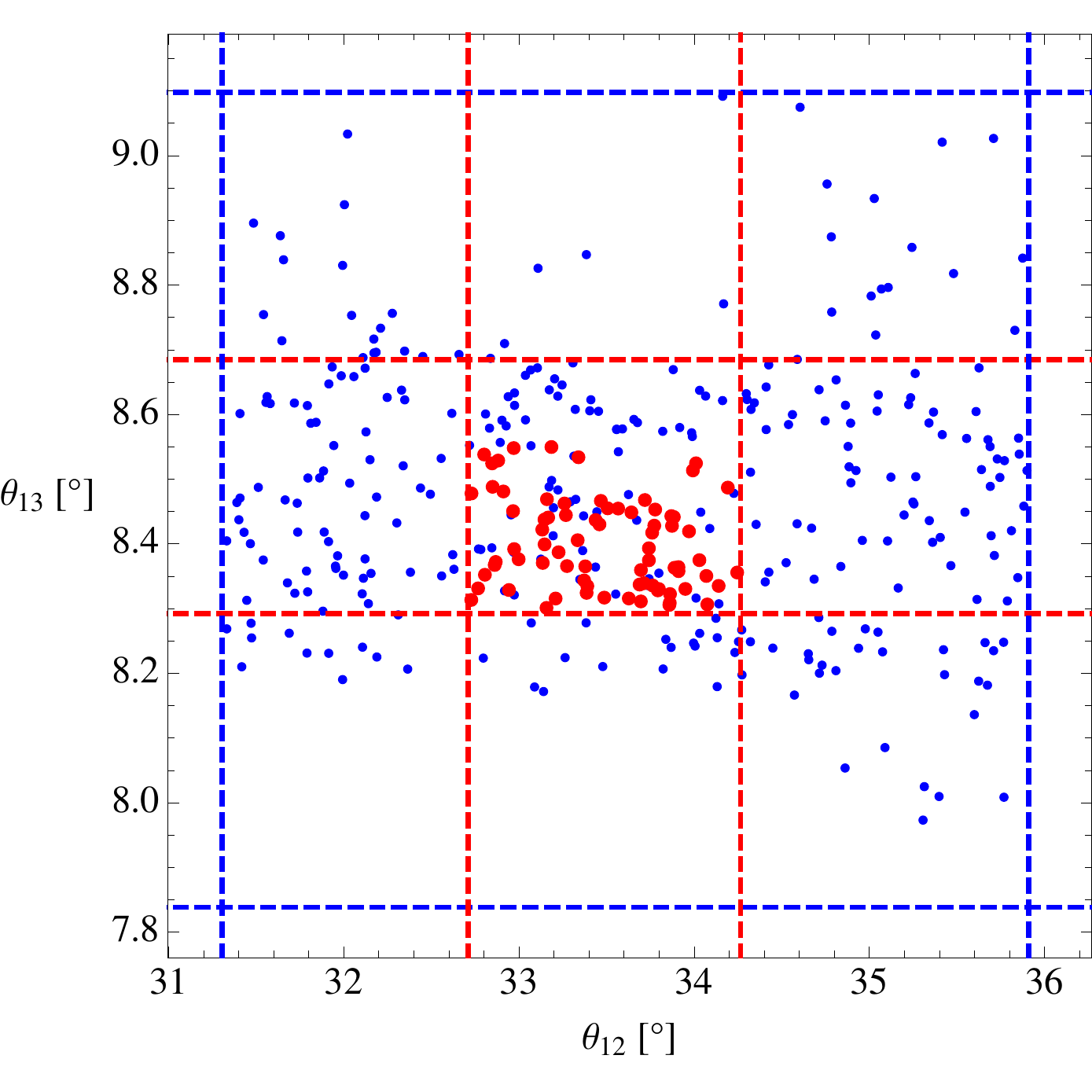}
\includegraphics[scale=0.49]{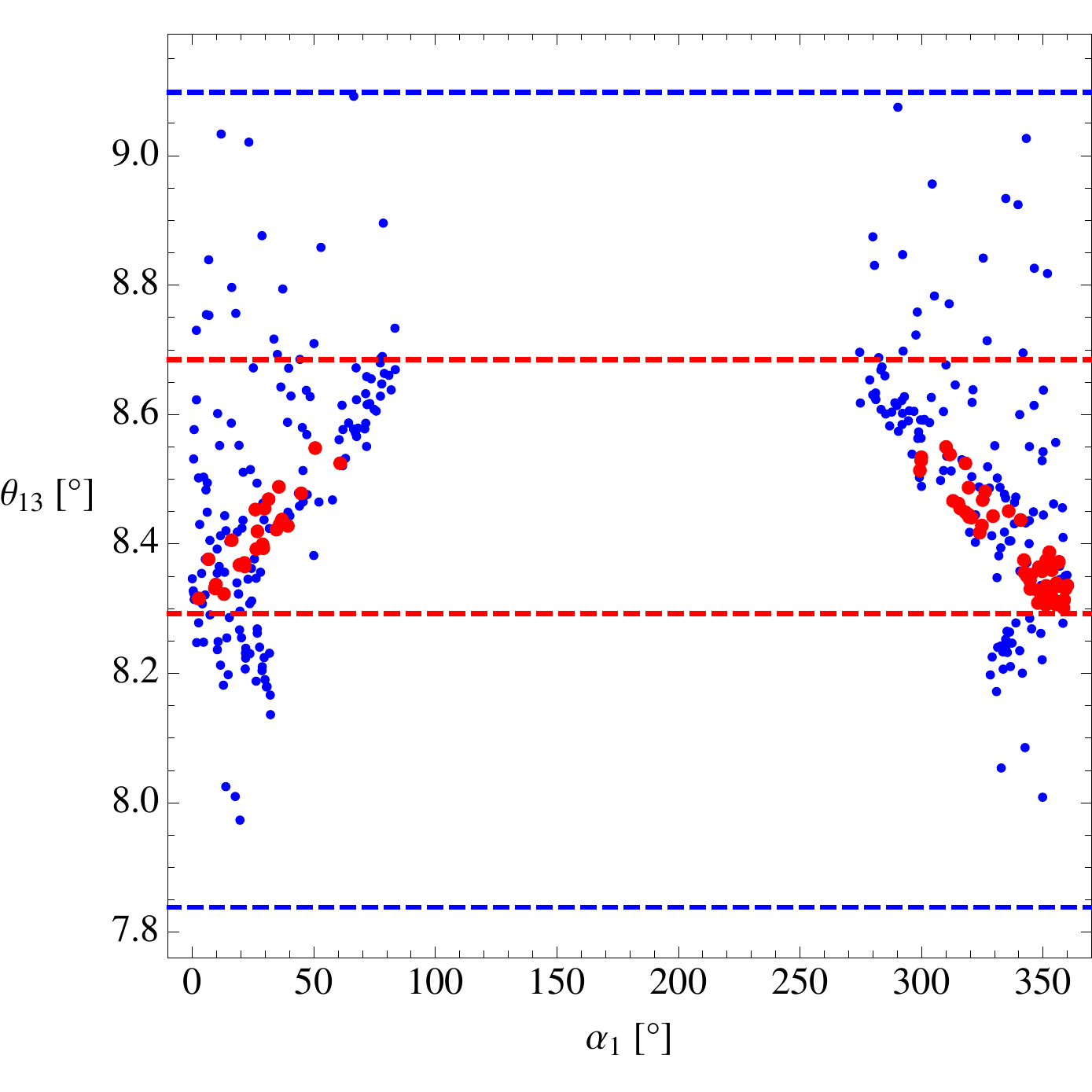} \hspace{0.7cm}
\includegraphics[scale=0.49]{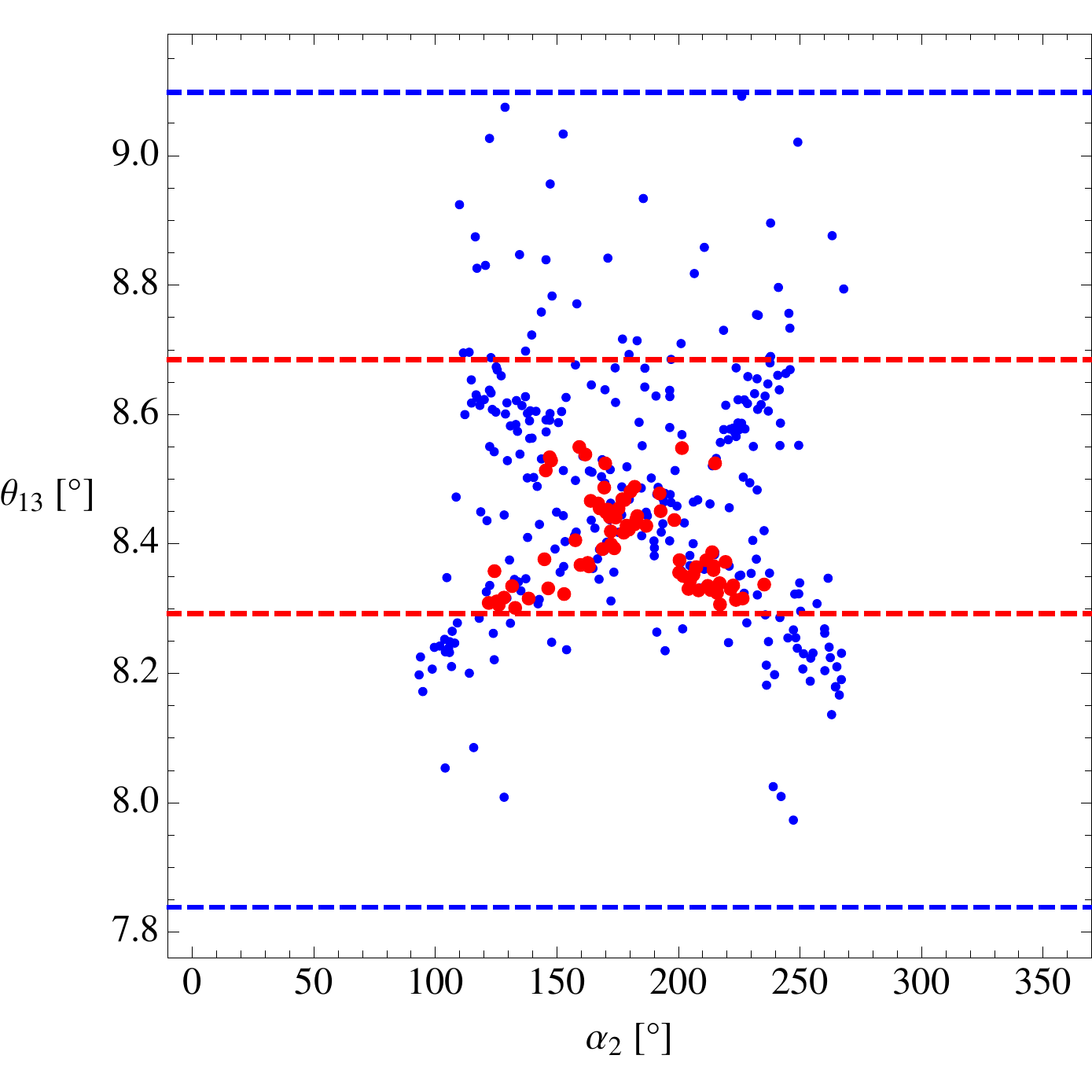}
\includegraphics[scale=0.49]{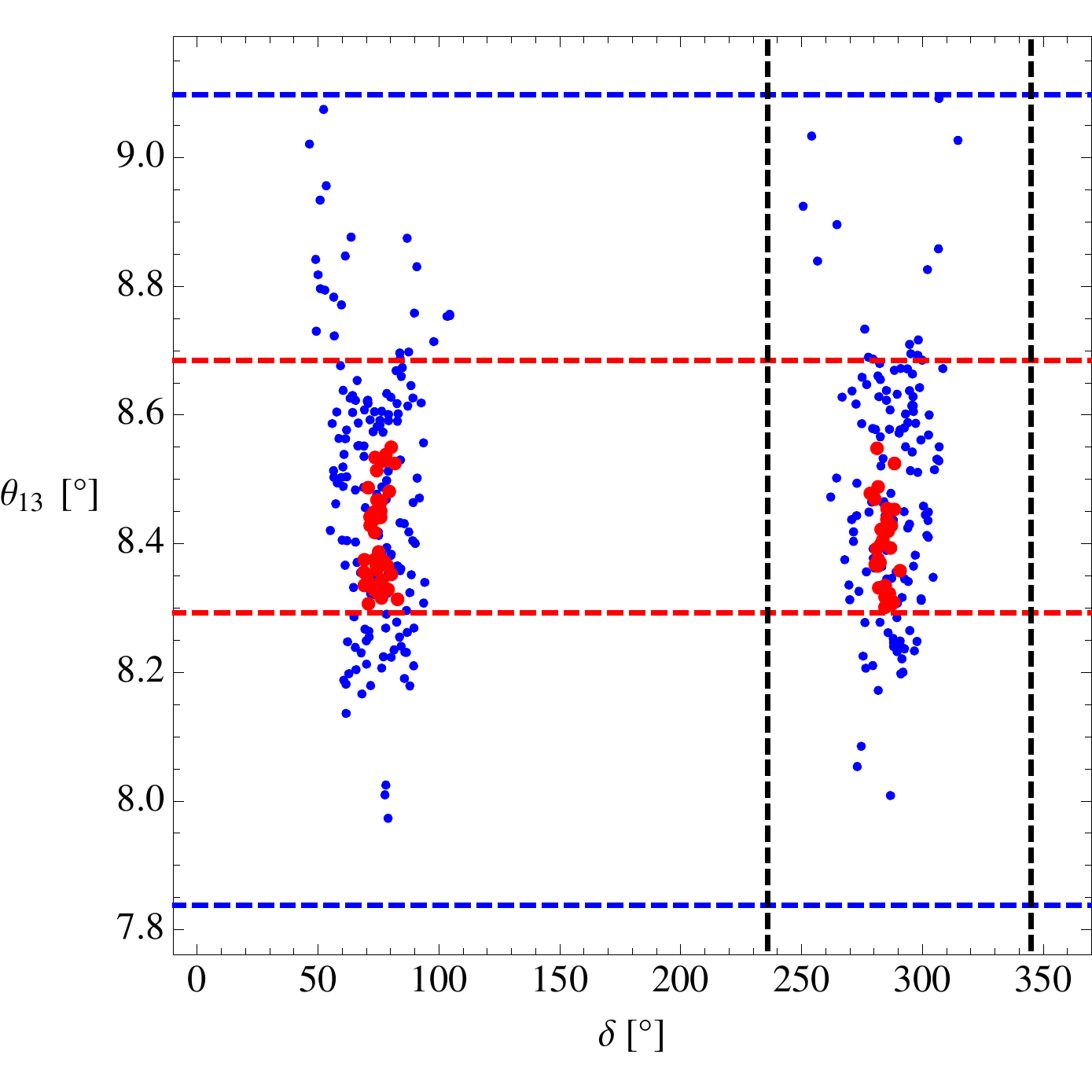} \hspace{0.7cm}
\includegraphics[scale=0.49]{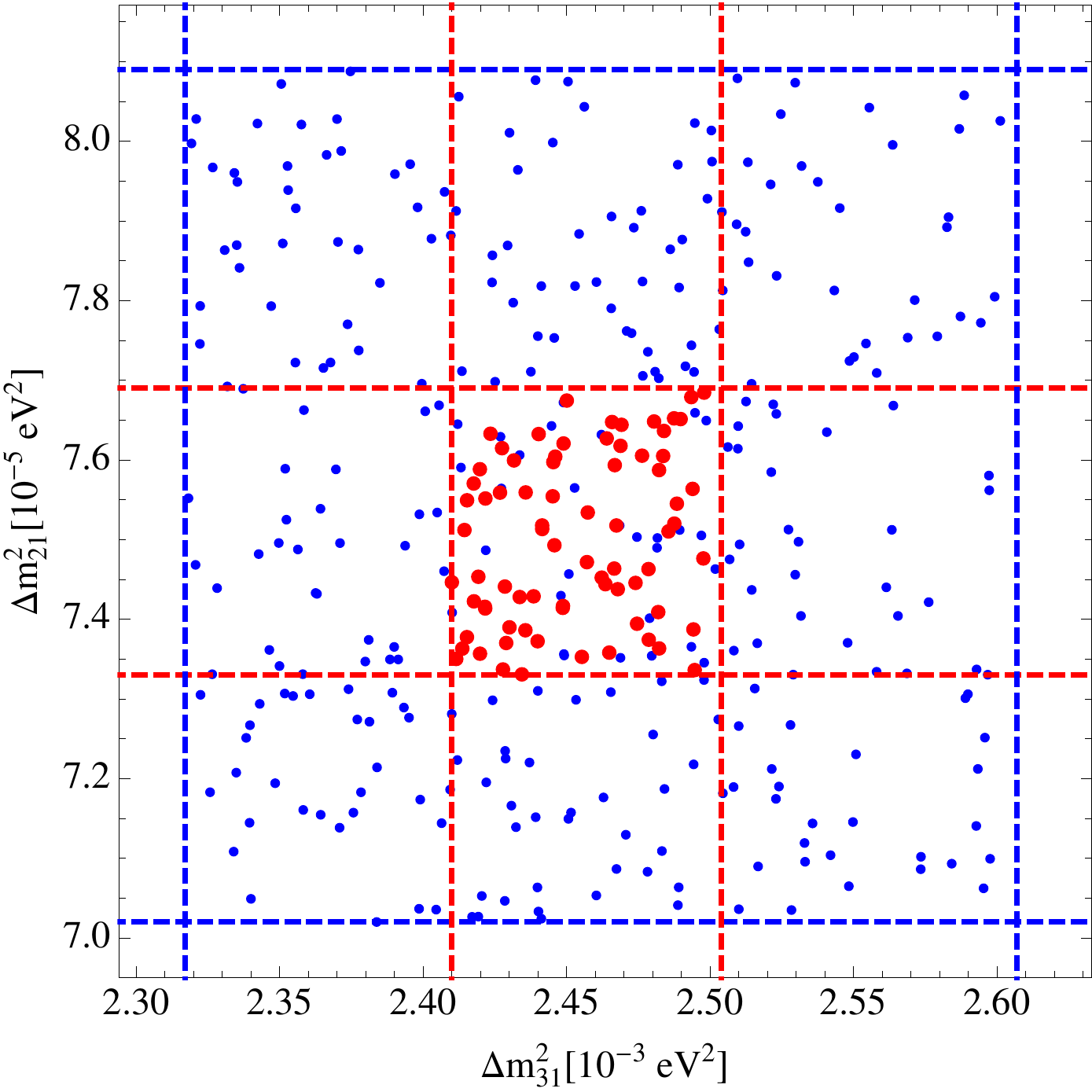}
\caption{
Results of our numerical parameter scan.
Blue (red) points are in agreement within 3$\sigma$ (1$\sigma$) of the low energy
neutrino masses and mixings and $Y_B$ in our model.
The allowed experimental 3$\sigma$ (1$\sigma$) regions are
limited by blue (red) dashed lines. The black dashed lines represent the 1$\sigma$
range for the not directly measured CP phase $\delta$ from the global
fit \cite{Gonzalez-Garcia:2014bfa}.
}
\label{fig:mixing_parameters}
\end{figure}

Before we come to our results for the normal ordering we want to
comment briefly on the inverted ordering. In our numerical scan
we were not able to find any points in agreement within 3$\sigma$
with all the mentioned observables. We restricted $c \leq 0.2$
and neglected points where due to a fine-tuned cancellation
the NLO corrections were artificially enhanced.
Hence, we conclude that this ordering
is still excluded like in the original model.

The results of our scan for the masses and mixing angles is shown in
Fig.~\ref{fig:mixing_parameters} where the careful reader might note
first that now we have as well found parameter points that are in agreement
within 1$\sigma$ with all observables. That seems to be surprising since
we have added here an additional constraint and apart from this expect rather
small deviations from the original model. But there are two things coming together: First of all,
due to the correction we can now allow for smaller values of $\theta_{23}^{\text{PMNS}}$
down to about $44^\circ$ and furthermore we use here the updated results
from the nu-fit collaboration \cite{Gonzalez-Garcia:2014bfa} which allows
for $\theta_{23}^{\text{PMNS}} = 45^\circ$ even at 1$\sigma$.

\begin{figure}
\centering
\includegraphics[scale=0.55]{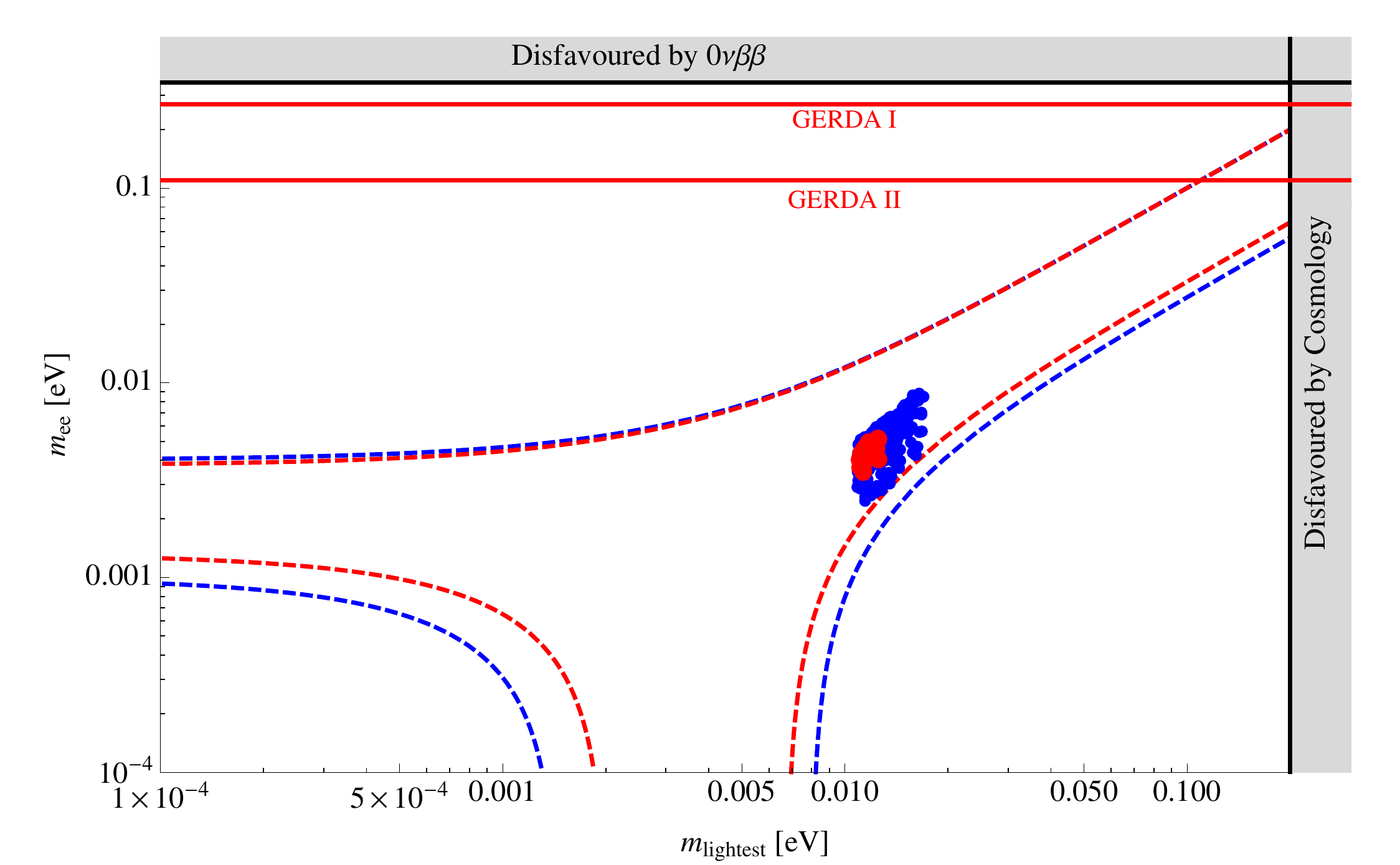}
\caption{
Prediction for the effective neutrino mass $m_{ee}$ accessible in neutrinoless double beta decay
experiments as a function of the lightest neutrino mass $m_{1}$.
The allowed experimental 3$\sigma$ (1$\sigma$) regions for the masses and mixing angles
in the case of normal ordering are limited by blue (red) dashed lines.
Blue (red) points are in agreement within 3$\sigma$ (1$\sigma$) of the low energy
neutrino masses and mixings and $Y_B$ in our model.
The grey region on the right side shows the bounds on the
lightest mass from cosmology \cite{Ade:2013zuv} and the grey region in the upper part displays the upper
bound on the effective mass from the EXO experiment \cite{Albert:2014awa}.
The red, straight lines represent the sensitivity of GERDA phase I respectively GERDA phase II \cite{Smolnikov:2008fu}.
}
\label{fig:mee_plot}
\end{figure}

The second thing to note is that now the correlations between $\theta_{13}^{\text{PMNS}}$
and the phases is much weaker which can be explained by the fact
that now we have on top another complex parameter in the game.
But still the phases are not in arbitrary ranges but we find
\begin{align}
\delta &\in [47 ^{\circ}, 104^{\circ}] ~\text{or}~ [250 ^{\circ}, 316^{\circ}] \;, \\
\alpha_1 &\in [0^{\circ}, 85^{\circ}] ~\text{or}~ [275^{\circ}, 360^{\circ}] \;, \\
\alpha_2 &\in [94 ^{\circ}, 269^{\circ}] \;,
\end{align}
For the Jarlskog invariant which determines the CP violation in neutrino oscillations we find values between $\pm$($0.026 - 0.035$).
The restricted ranges for the phases imply of course also restrictions
on the predictions for neutrinoless double beta decay, see Fig.~\ref{fig:mee_plot}.
But more restrictive in this case is nevertheless the constraint on the mass scale
where the lower bound is mostly determined by the mass sum rule. We obtain for the lightest neutrino mass $m_1$ values between 10.5 meV to 17.6 meV. In fact, our prediction for $m_{ee}$ is rather precise to be in the narrow
range from 2.3~meV to 9.2~meV. This is way below the sensitivity of any experiment in the near future so
that any evidence for neutrinoless double beta decay would rule out this model.

Related to the mass scale are as well two other observables. First of all there is the sum
of the neutrino masses
\begin{equation}
\sum m_{\nu} \in (0.074 - 0.089) ~\text{eV} \;,
\end{equation}
which might be determined from cosmology. So far
there is only an upper bound  \cite{Ade:2013zuv}
\begin{equation}
\sum m_{\nu} < 0.23 \text{ eV,}
\end{equation} 
which is well in agreement with our prediction.
The second observable is the kinematic mass $m_{\beta}$  as measured
in the KATRIN experiment \cite{Angrik:2005ep}
which is given as
\begin{equation}
m_{\beta}^{2}=m_{1}^{2}c_{12}^{2}c_{13}^{2}+m_{2}^{2}s_{12}^{2}c_{13}^{2}+m_{3}^{2}s_{13}^{2} \;.
\label{eq:katrin}
\end{equation}
Here we predict $m_{\beta} \approx (0.014-0.019)$~eV which is again way below the
projected reach of $m_{\beta} > 0.2$~eV.

\subsection{Leptogenesis}

In this section we show the results of our parameter scan
relevant for leptogenesis where we have implemented the
formulas given in section~\ref{sec:AnalyticalLeptogenesis} to
calculate the generated baryon asymmetry. 

\begin{figure}
\centering
\includegraphics[scale=0.4]{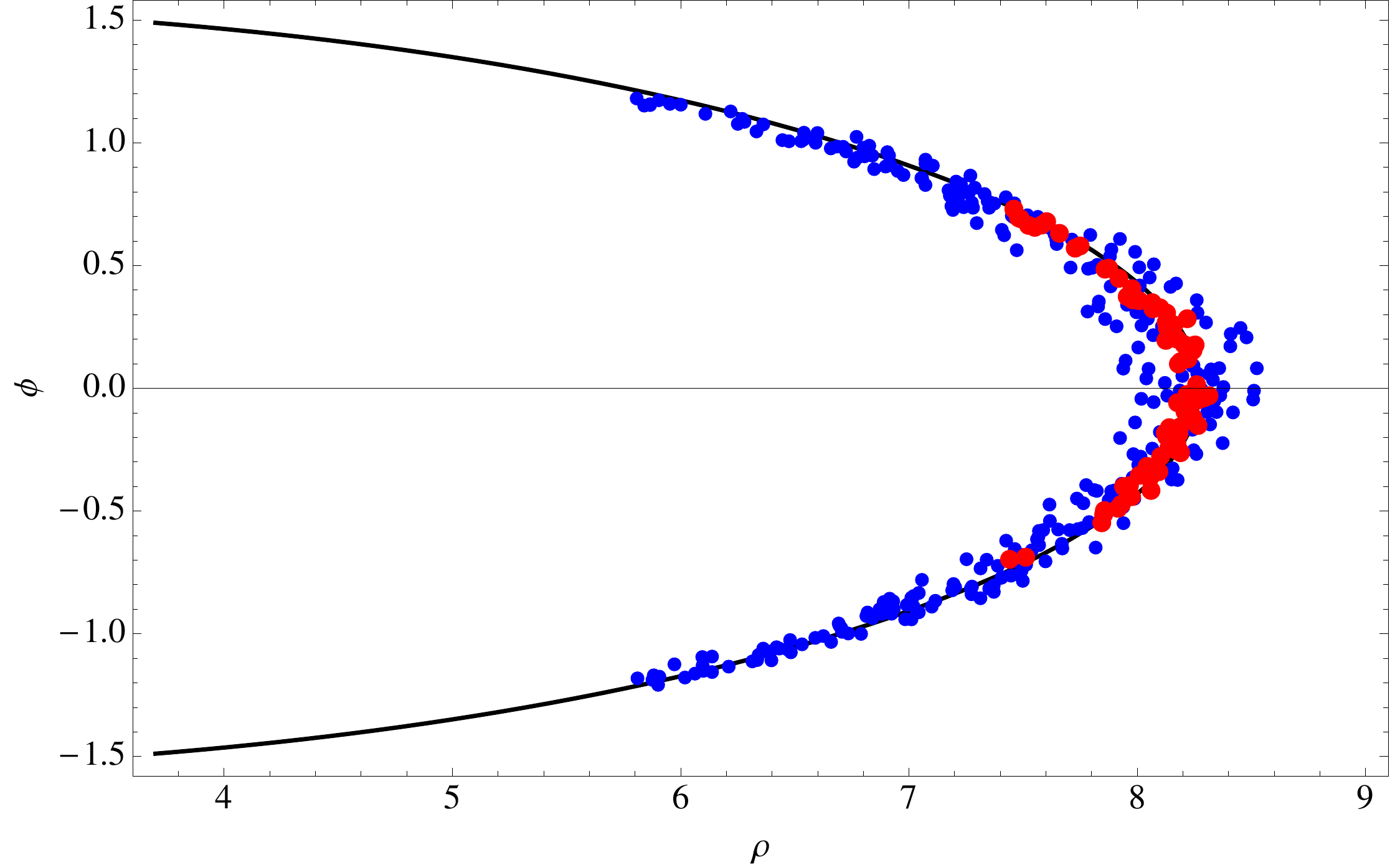}
\caption{
The relation between $\phi$ and $\rho$ according to eq.~\eqref{alphaphi}
(black line), cf.~Fig.~\ref{fig:PhiAlpha}, and blue (red) points from our numerical scan in agreement
within 3$\sigma$ (1$\sigma$) of the low energy
neutrino masses and mixings and $Y_B$.
}
\label{fig:PhiAlphaCompare}
\end{figure}

Before we actually discuss the results for the asymmetry itself
we first want to note that the results from our analytical estimates
are quite good. For instance, in Fig.~\ref{fig:PhiAlphaCompare}
we show the relation between $\phi$ and $\rho$ from eq.~\eqref{alphaphi}
and from our numerical scan. The agreement is striking although in the analytical
estimates we have neglected for instance RGE effects which are nevertheless
not very large in the allowed mass range. The biggest difference is in the allowed
range for $\rho$. To avoid the resonance condition we only demanded $\rho \gtrsim 3.7$
while we find here $\rho \gtrsim 5.8$. But here not only the ratio of the mass squared
differences enter, but the two values of the mass squared differences independently.

\begin{figure}
\centering
\includegraphics[scale=0.49]{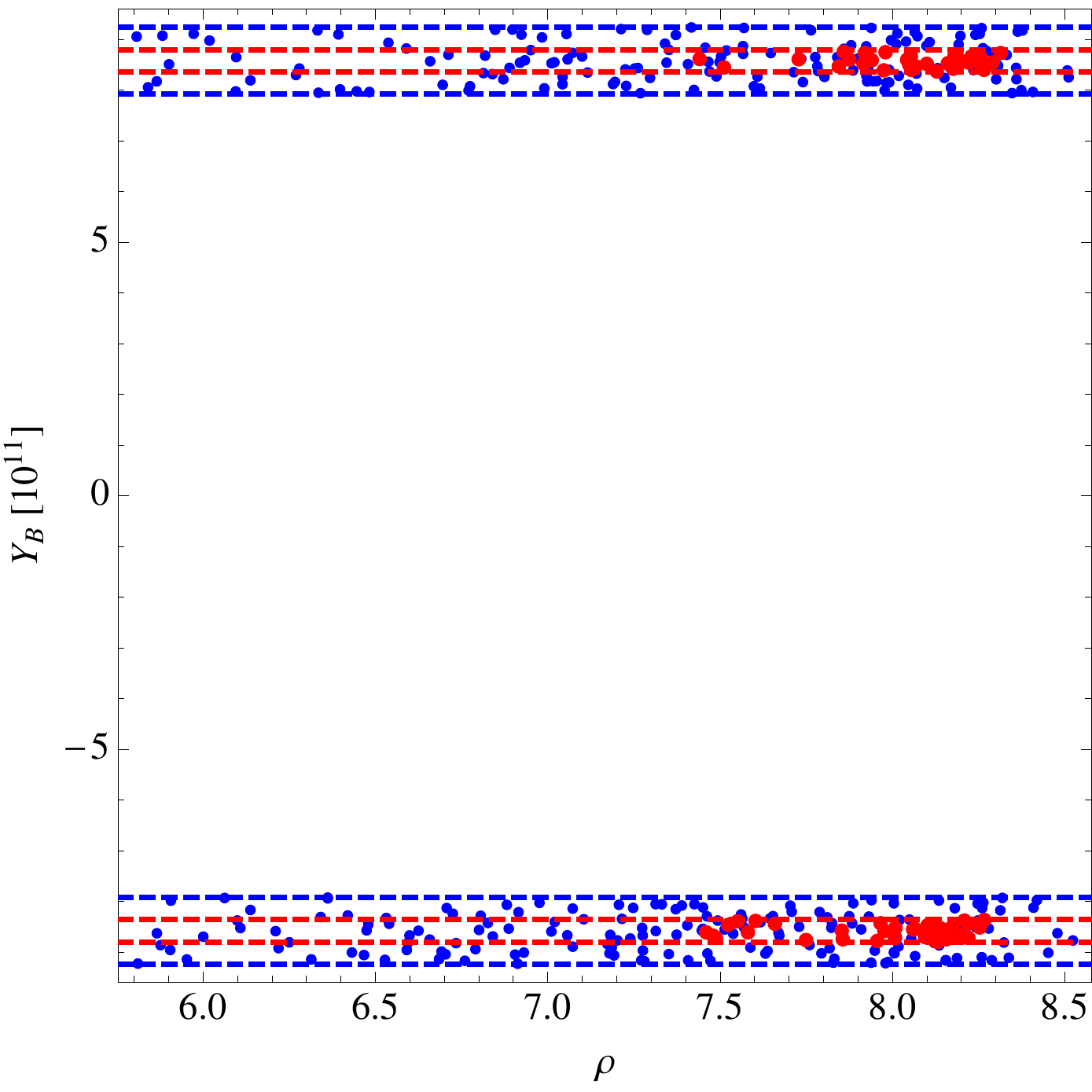} \hspace{0.7cm}
\includegraphics[scale=0.49]{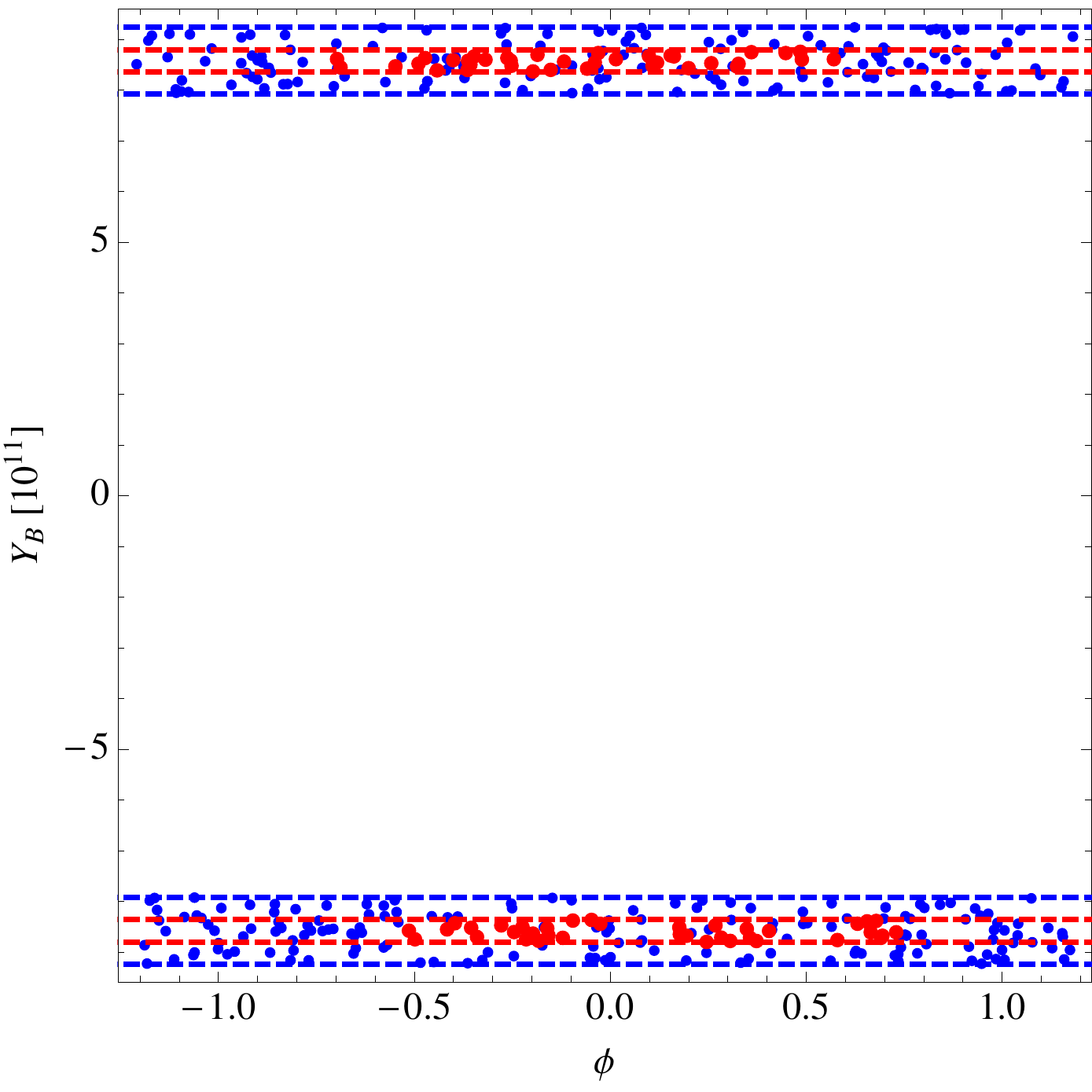} \\
\includegraphics[scale=0.49]{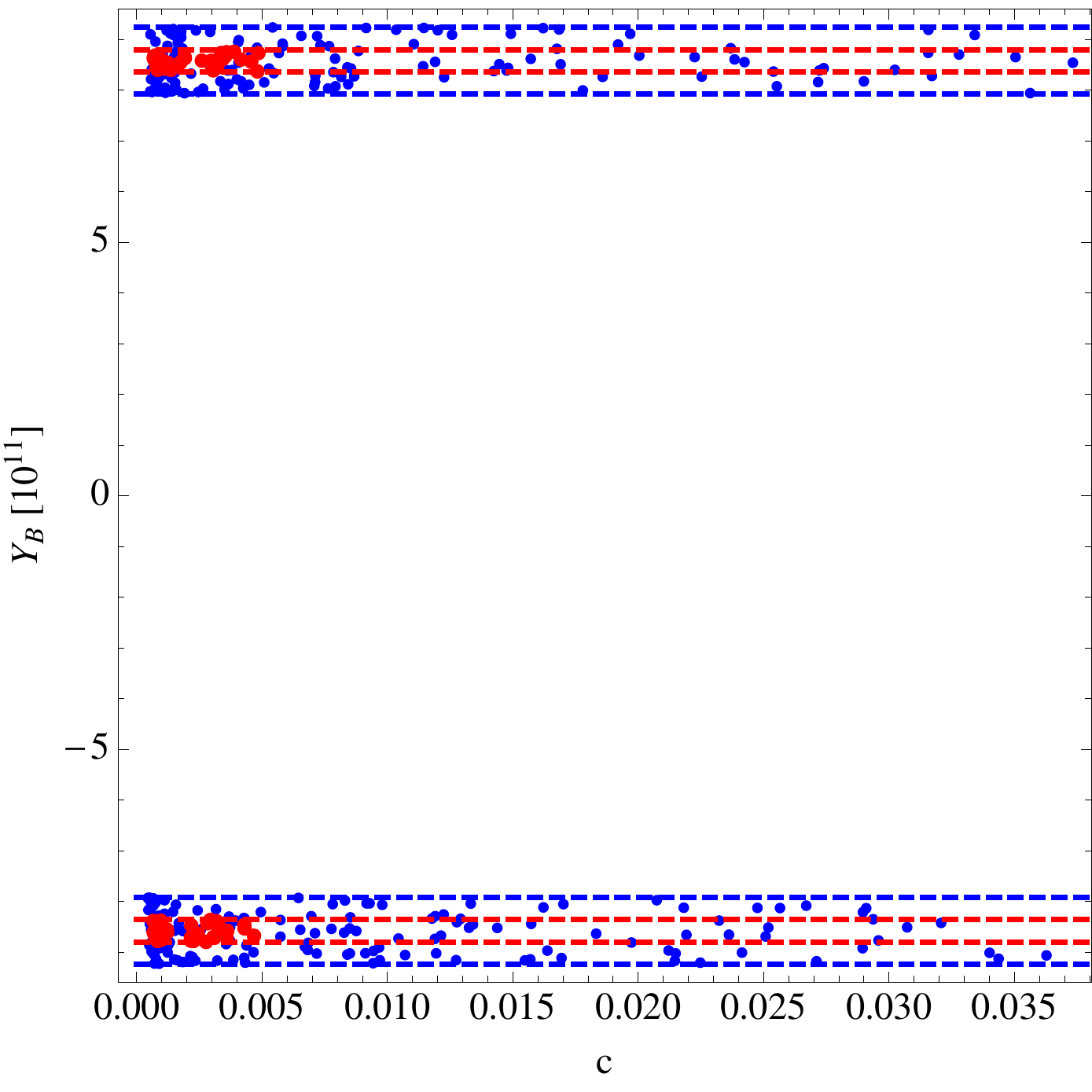} \hspace{0.7cm}
\includegraphics[scale=0.49]{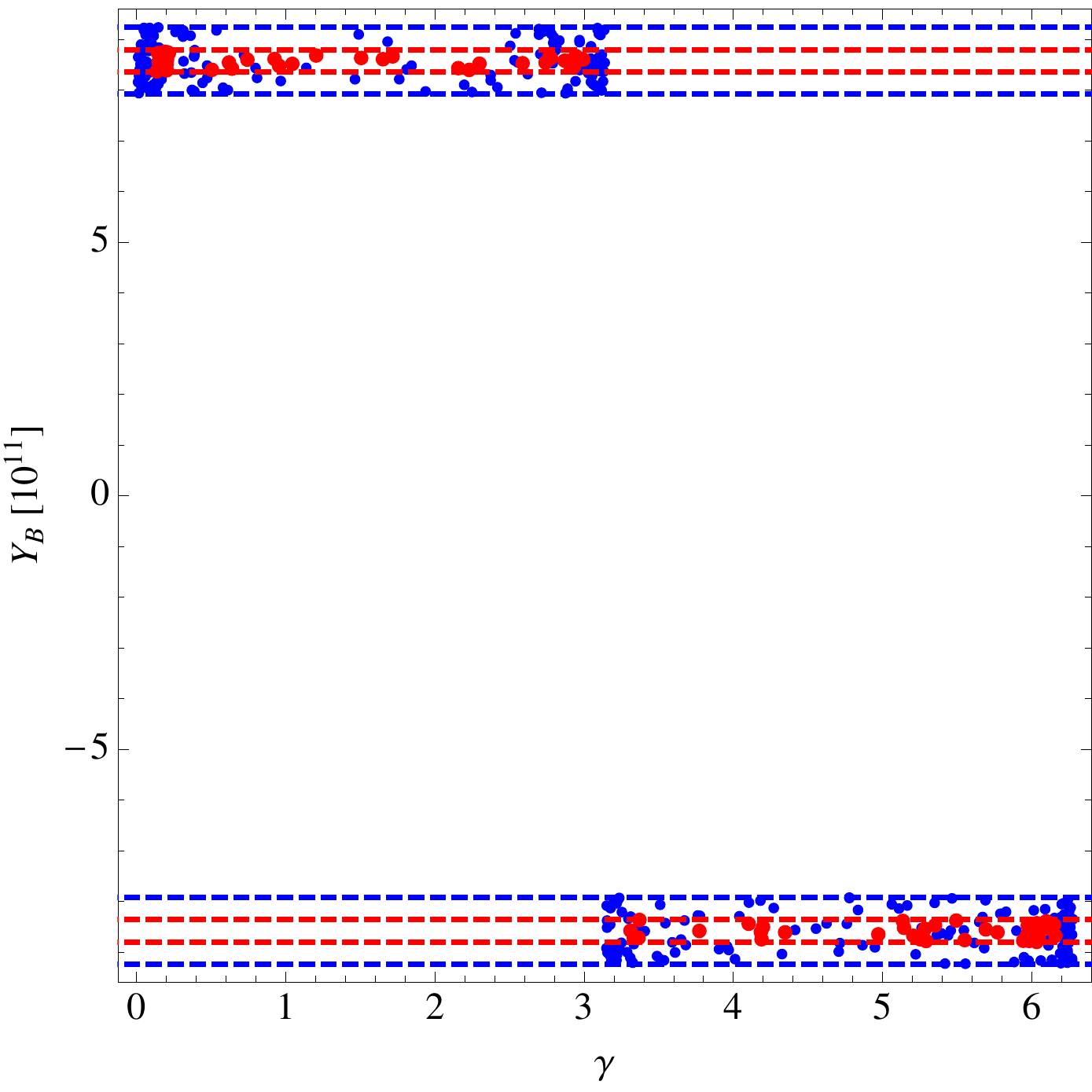} \\
\caption{
Results of our numerical scan for the total baryon asymmetry $Y_B$
in dependence of the four most relevant parameters.
Blue (red) points are in agreement within 3$\sigma$ (1$\sigma$) of the low energy
neutrino masses and mixings and $Y_B$ in our model.
}
\label{fig:YB_num}
\end{figure}

Now that we are convinced that our analytical estimates have been good
we discuss the dependence of $Y_B$ on the four most relevant parameters
as discussed in section~\ref{sec:AnalyticalLeptogenesis}. The biggest advantage
of our numerical scan over the analytical estimates is that it allows us to
use all available data on neutrino masses and mixing to constrain the allowed
parameter space.

We have already seen in Fig.~\ref{fig:PhiAlphaCompare} 
that the values of $\rho$ get constrained which is again visible in Fig.~\ref{fig:YB_num}.
While a priori we only knew that $\rho \gtrsim 3.7$ and less than about 9 we now
see that only the range from $5.8$ to $8.5$ is allowed ($7.4 - 8.3$ at 1$\sigma$).
And since $\rho$ and $\phi$ are not independent but related via eq.~\eqref{alphaphi},
the phase $\phi$ gets constrained as well to the range $[-1.2;1.2]$ ($[-0.5;0.7]$ at
1$\sigma$).

The new parameters $c$ and $\gamma$ are nevertheless more interesting than
$\rho$ and $\phi$ which are mostly constrained by the
neutrino masses and for which we would have found similar results
already in the previous model. In Fig.~\ref{fig:YB_num} we have shown
the dependence of $Y_B$ on this new parameters.

The first thing to note, is that $c$ is indeed a small parameter in the range
from $0.7 \cdot 10^{-3}$ to $4.9 \cdot 10^{-3}$. From the
model building point of view such a small value is justified. Remember
that the leading order Yukawa coupling is a dimension three
operator in the
superpotential while the correction proportional to $c$ is coming from a
dimension seven operator.
Also note that alone from a constraint on $Y_B$
$c$ could have been much larger or smaller depending of course on the
value of $\gamma$ and the other
parameters. This is different here because the mixing angles get corrections
of order $c$ and this implies the constraint shown here.

Finally, note that the allowed range for $\gamma$ is only weakly constrained.
Nevertheless, it is interesting that the sign of $Y_B$ is completely determined by $\gamma$.
This is somewhat surprising because in our estimates from section~\ref{sec:AnalyticalLeptogenesis}
the sign of $\epsilon^\tau$ depends on other parameters which could induce a sign flip,
which can be seen for instance in the upper plot of Fig.~\ref{fig:y_t}. But
after applying all experimental constraints the correlation is striking.

Combined with the analytic analysis, we see that this correlation is a result 
 of the fact that $Y_B$ is dominated by $Y_1$, which is again dominated 
 by the first term in $\epsilon_1^\tau$, where $\cos\phi_2$ is negative 
and $f(m_1/m_2)$ is positive. Neglecting the subdominant terms, 
we have $Y_B\propto\sin\gamma$. 
The analytical estimates for the efficiency factors we
are using provide results with an estimated precision of (20-30)$\%$ compared
to the full numerical results solving the Boltzmann equations.
This is more than sufficient for the purposes of our study.
The other predictions for the light neutrino masses and mixing parameters
would only mildly change because they are mostly governed by the leading order
values (a 30\% correction to $c$ would have only little impact on them).
It is also worth mentioning that 
the complex Yukawa and the Majorana phases are both necessary 
CP-violating sources to generate a successful baryon asymmetry 
via leptogenesis while in accordance with all the low energy 
constraints. It is also noticeable that $Y_B$ would be 
strongly suppressed if $\cos\phi_2 \cong 0$. As it 
follows from Fig.~\ref{fig:PhiAlpha}, values of  $\cos\phi_2 \cong 0$
are excluded in the model we are considering since
$\cos\phi_2$ can have values only in a narrow interval
around $-1$, namely, $(-1, -0.95)$ for $\rho\gtrsim 5.8$, 
which also means that the Majorana phases contribute maximally
to the asymmetry. In order to investigate the role of the Dirac 
phase we need a different parametrisation of the neutrino 
Yukawa coupling to see the relation explicitly, which is beyond the 
scope of the current work.

\section{Summary and Conclusions}

In this paper we have revised the SU(5)~$\times$~A$_5$ golden ratio 
GUT flavour model from \cite{Gehrlein:2014wda}
with the aim to include as well successful leptogenesis. In the 
original setup this was
not possible. As it turns out we only have to add two 
additional pairs of messenger
fields but no additional symmetries or flavon fields to do this. 
We find that this induces a small correction to the neutrino Yukawa matrix, 
which can generate a sizeable baryon asymmetry,
but as well implies some modifications for the predictions 
of the masses and mixing angles
of the original model. In an extensive numerical scan 
we could show that we can simultaneously
accommodate successfully the observed neutrino masses, 
mixing angles and possibly baryon asymmetry.
And even more our setup is so constrained that we 
predict several correlations or ranges for
observables yet to be measured.

One of the most striking features of our original model - 
the sum rule for the neutrino 
masses -  remains valid up to a insignificant
correction. From this we can again derive a lower
bound for the lightest neutrino masses $m_1 \gtrsim 0.011$ eV 
and rule out 
the neutrino mass spectrum with inverted ordering. 
This is already a very strong prediction.

Due to the additional complex parameter and the additional 
constraint on $Y_B$
the allowed ranges for $\alpha_2$ has 
shrunk from $ 70^{\circ}-290^{\circ}$ in the original model
to $94^{\circ}-269^{\circ}$. Whereas the allowed 
regions for $\alpha_1$ and $\delta$ remain similar
compared to the original model. Namely, we find 
now $\alpha_1$ to be in $0^{\circ}-85^{\circ}$ or
$275^{\circ}-360^{\circ}$ and $\delta$ to be in $47^{\circ}-105^{\circ}$ 
or $250^{\circ}-316^{\circ}$.
The strong correlation between $\theta_{13}$ and 
the Majorana phases is now weakened due to the additional
complex parameter we introduced
in the model. It is also important to note,
that we find here points which are in agreement 
within 1$\sigma$ with all neutrino observables.
This is due to the fact that we now allow for smaller 
values of $\theta_{23}$ but we also use
here the updated fit results from \cite{Gonzalez-Garcia:2014bfa} 
where maximal atmospheric mixing 
is again allowed at the 1$\sigma$ level. 
Nevertheless, a precise measurement of $\theta_{23}$ which
deviates significantly from maximal mixing can rule out the presented model.
Since we limit the allowed ranges for the CP violating 
phases and the light neutrino masses we predict
as well the effective Majorana mass observable in neutrinoless double beta
decay to be in the narrow range ($2.3 - 9.2$)~meV.
This is beyond the reach of ongoing experiments and upcoming experiments
which will begin taking data in the near future, 
but it will be certainly tested in the future.

For the baryon asymmetry $Y_B$ we find
 in the approximation used to calculate it 
good agreement with the most recent data and this is done
by only introducing one additional operator which involves 
one new complex parameter with a modulus $c$ having a value
in the range $0.7 \cdot 10^{-3}$ to $4.9 \cdot 10^{-3}$. 
The phase of this additional parameter
at the 3$\sigma$ level is not much constrained but it governs
the sign of $Y_B$.

What we did not discuss in the present article 
is that some of the features of the original model, 
like the Yukawa coupling ratios
$y_\tau/y_b \approx - 3/2$, remain valid in the modified model
implying non-trivial constraints on the
spectrum of the supersymmetric partners of the Standard Model particles.

In summary we have succeeded to modify the model from 
\cite{Gehrlein:2014wda} to include
viable leptogenesis by only introducing a minimal correction. 
The model presented here is, to our knowledge, the 
first GUT $\text{A}_5$ golden ratio flavour model 
in which it is possible to have successful leptogenesis. 
All observables lie within the
measured ranges and for the not yet measured quantities
in the neutrino sector 
(the type of the neutrino mass spectrum, 
the absolute scale and the sum of the neutrino masses, the effective 
Majorana mass in neutrinoless double beta decay, the CP 
violation phases in the PMNS matrix),
we make predictions. 
An appealing feature of the model is its rather small number 
of parameters, which makes the model 
very predictive and testable.

\section*{Acknowledgements}
The work of X.~Zhang was done during her visit of SISSA, 
and it was supported by the graduate school of Peking University (grant
number
zzsq2014000091). 
X.~Zhang would like to thank Prof. Petcov 
for hospitality at SISSA, I.~Girardi, A.~Titov and 
A.~J.~Stuart for discussions, and A.~J.~Stuart for sharing 
his code on $A_5$ contractions. S.T.P. acknowledges very 
useful discussions with E. Molinaro.                        
This work was supported in part  by the European Union FP7
ITN INVISIBLES (Marie Curie Actions, PITN-GA-2011-289442-INVISIBLES),
by the INFN program on Theoretical Astroparticle Physics (TASP),
by the research grant  2012CPPYP7 ({\sl  Theoretical Astroparticle Physics})
under the program  PRIN 2012 funded by the 
Italian MIUR and by the World Premier International Research Center
Initiative (WPI Initiative), MEXT, Japan (STP).

\end{document}